\documentclass{article} 
\usepackage{aaspp4}
\usepackage{psfig}

\newcommand\fx{F_{\rm X}}
\newcommand\tskin{\tau_s}
\newcommand\fr{F_{\rm rep}}
\newcommand\msun{{\,M_\odot}}

\def\lsim{\lower.5ex\hbox{$\; \buildrel < \over \sim \;$}}
\def\gsim{\lower.5ex\hbox{$\; \buildrel > \over \sim \;$}}

\def\t{\ifmmode {\tau} \else $\tau$ \fi}

\def\sw{Schwarzschild~}
\def\cm{\ifmmode {\rm cm}^{-1} \else cm$^{-1}$ \fi}
\def\s{\ifmmode {\rm s}^{-1} \else s$^{-1}$ \fi}
\def\cc{\ifmmode {\rm cm}^{-3} \else cm$^{-3}$ \fi}
\def\cs{\ifmmode {\rm cm}^{-2} \else cm$^{-2}$ \fi}
\def\g{\ifmmode \gamma \else $\gamma$\fi}
\def\l{\ifmmode \lambda \else $\lambda$\fi}
\def\ls{$\lambda$~}

\def\t{\ifmmode \tau \else $\tau$\fi}
\def\G{\ifmmode \Gamma \else $\Gamma$\fi}
\def\Gt{\ifmmode \tilde{\Gamma} \else $\tilde{\Gamma}$\fi}

\def\kms{\ifmmode {\rm km\ s}^{-1} \else km s$^{-1}$\fi}

\def\fe{Fe K$\alpha$~}

\begin{document}

\title{Modeling the X-ray -- UV Correlations in NGC 3516 }

\author{Demosthenes Kazanas\altaffilmark{1}\& Sergei Nayakshin\altaffilmark{1,2}}

\altaffiltext{1}{LHEA, NASA/GSFC Code 661, Greenbelt, MD 20771}
\altaffiltext{2}{NAS/NRC Research Associate}

\font\rom=cmr10
\centerline{\rom submitted to Astrophys. J.}

\begin{abstract}
\noindent We test the ``reprocessing paradigm" of the optical -- UV
AGN variability, according to which the variations in this wavelength
range are driven by a variable X-ray component, by detailed modeling
of the correlated X-ray -- optical (\l\ls 3590 and 5510 \AA)
variability of the recent multiwavelength campaign of NGC 3516. To
this end we produce model optical light curves by convolving the
observed X-ray flux with the response function of an infinite, thin
accretion disk, illuminated by a point-like X-ray source at a given
height $h_X$ above the compact object (the lamp-post model) and
compare their properties (amplitude, morphology, lags) to those
observed. Special attention is given to the correct computation of the
X-ray albedo of the disk, by including an X-ray heated ionized layer
in hydrostatic equilibrium on its surface. We further compute the
X-ray reflection response at two energies ($E = 1,\, 20$ keV) and
argue for the possibility of hard lags in their cross spectra. We also
compute the continuum Optical -- UV and the X-ray reflection spectra
as well as the \fe fluorescent line profiles which we also compare to
observations. Despite the large ($\simeq 50\%$) amplitude excursions
of the X-ray flux, the model optical light curves exhibit variability
amplitudes of 3 -- 4 \%, not unlike those observed. The model light
curves exhibit clearly a feature associated with a large X-ray flux
excursion not seen in the data, arguing for a reprocessing region of
size $\gsim 10^{15}$ cm, even though no lags are discernible in the
cross correlation functions (CCF) of the \ls 3590 and \ls 5510 \AA~
model light curves. However, the (CCF) between the X-ray and the model
optical variations show clear lags of 0.1 and 0.25 days for black hole
masses $M = 10^7, \, 10^8$ M$_{\odot}$ respectively, not apparent in
the data.  The synchrony of X-ray - optical variations points toward
the smaller mass values ($M = 10^7$ M$_{\odot}$), which however are
inconsistent with the model X-ray reflection spectra which exhibit
several features below $\sim$ 4 keV and above $\simeq 9$ keV not
observed in the data. In fact, the model X-ray reflection spectra are
roughly consistent with observations only for $M \gsim 10^8$
M$_{\odot}$. The observed \fe line profiles are too broad to be
consistent with our models. Lowering the distance of the X-ray source
from the disk, $h_X$, helps a little in this respect, but only at the
expense of reducing unrealistically the amplitude of Optical--UV
variations.  Our conclusion is that the combination of the observed
optical/UV/X-ray spectral and timing observations are inconsistent
with the lamp-post model geometry for NGC~3516.

\end{abstract}


\section{Introduction}



A great deal of effort and observing time has been expended in the
past decade or so in a systematic effort to ``map" the central regions
of Active Galactic Nuclei (hereafter AGN). Because the relevant sizes
are too small to image with current technology, the attempted
``mapping" has been effected through the time reverberation technique,
i.e., the monitoring of the system in response to fluctuations in
luminosity across the electromagnetic spectrum and the comparison of
the relative amplitudes and lags between different wavelengths to
models.

While the original reverberation effort was aimed at estimating 
the size of the broad line region (BLR) by  measuring the lags
between variations in the ionizing continuum and the broad emission
lines (see Netzer \& Peterson 1997 for a review), it was quickly 
realized that the same data could also be used to test models of 
the continuum emission itself  by measuring the lags between 
variations in the UV and optical flux. 
The AGN optical -- UV (hereafter O -- UV) continuum 
is generally dominated by a broad quasi-thermal component,
the so-called Big Blue Bump (hereafter BBB), thought to be due to -- 
and modeled as -- the emission of a geometrically thin, optically 
thick accretion disk radiating in black body form the locally 
dissipated accretion kinetic energy (Malkan \& Sargent 1983; 
Malkan 1984; Laor \& Netzer 1989; Sun \& Malkan 1989). 
Thus, monitoring the O -- UV continuum variability and the corresponding 
interband lags, places constraints on the models of such accretion disks. 

It became apparent early on in this extensive mapping effort that the 
interband lags between optical and UV wavelengths were far shorter 
than those expected by most modes of information propagation in a 
thin accretion disk (the lags of the earlier campaigns
were generally shorter than the sampling rates). 
This situation prompted the suggestion that the correlated  
O -- UV continuum variability is due entirely to reprocessing 
of the more variable X-ray component by the 
geometrically thin accretion disk responsible for the emission 
associated with the BBB feature (Krolik et al 1991). 
This suggestion appeared to also fit nicely with the evidence 
of X-ray reflection by cold matter in AGN (Pounds et al. 1990) 
and the presence of relativistically broadened \fe lines 
(e.g., Tanaka et al. 1995; see also a recent review by Fabian 
et al. 2000).

Thus, a ``picture" of the innermost regions of accreting black
holes began to emerge,  consisting of a geometrically thin, optically
thick accretion disk radiating away the dissipated kinetic
energy in black body form, supplemented by an X-ray source 
located at a height $h_X$ of a few \sw radii
above the disk (or modeled as a spherical source occupying
the inner part of the disk). The disk's soft photons were then
thought to be the seed photons that in interactions with the electrons 
of the hot corona produce the observed X-rays, which are
in turn reprocessed by the disk to provide the X-ray 
reflection feature, the \fe line, as well as the observed rapid, 
inter band variability between the optical and UV wavelengths.

While fits to the observed X-ray spectra within the general framework 
of this arrangement seem to be in good agreement with 
observations, as well known, spectra generally provide
information about column densities and optical depths rather
than about densities and actual lengths, which are the quantities 
needed to confirm the actual geometry of the X-ray -- UV -- O 
emission. Verification of the precise geometrical arrangement
requires a ``mapping" of this geometry through timing observations. 
Such a (so-called) reverberation mapping requires sufficiently 
high data sampling rate, set by the light crossing time from the 
X-ray source to the reprocessing disk. For a source size $\sim$ 
a few \sw radii of a $10^8$ M$_{\odot}$ black hole, this time scale 
(assuming that the source is located several  \sw radii above 
the disk) is of order of $10^3 - 10^4$ sec. At the  same time, 
the monitoring campaign has to be of sufficiently long duration 
to sample a large number of peaks and troughs
to produce a reliable interband cross correlation. 

There have been to date two monitoring campaigns specifically 
planned to comply with these 
criteria, namely those of NGC~7469 (Nandra et al. 1998, 2000)
and  of NGC~3516 (Edelson et al. 2000). Earlier coordinated 
multiwavelength observations, while forming a rather extended 
data base (see Nandra et al. 1998 for a summary), were 
deemed  of inadequate sampling rate to provide conclusive 
results. 
The results of the NGC 7469 campaign (Nandra et al. 1998) were
puzzling in that they indicated no clear correlation between
the X-ray variations and that of UV  emission, while the flux 
in both bands exhibited 50\% variations in amplitude. At the same time
there were clearly detected lags between the UV (\l 1315 \AA) and
the optical (6962 \AA) emission or order of 1--2 days 
(Collier et al. 1998). Revisiting these
observations, Nandra et al. (2000) have indicated that 
the softening of the X-ray spectrum with increasing UV flux
underestimated the soft X-ray emission. Taking this fact into
consideration, they showed that the long term ($\sim 10$ day) 
variations in the UV and X-ray bands were in fact ``in synch" 
with each other. However, the X-rays exhibited in addition 
strong variations on time scales of $\sim 10^4$ sec which had 
absolutely no counterpart in the UV or optical light curves.

Motivated by these results, Berkley, Kazanas \& Ozik (2000;
hereafter BKO), examined in detail these observations in 
the framework of the ``disk reprocessing model": They produced
simulated UV -- O light curves using the observed X-ray 
light curves as input in a geometric arrangement of a point
X-ray source at a given distance above the plane of a
disk, as thought to be the situation in AGN. Their conclusions
were that for any reasonable estimate of the black hole mass
($10^7 - 10^9$ M$_{\odot}$), both the UV  (\l 1315 \AA) and 
the optical (\l 6962 \AA)  emission should follow
closely that of the X-rays. The short time $(\tau \simeq 
10^4$ sec)  X-ray variability was always present in the model 
reprocessed emission, with almost undetectable ($\sim 0.1$
day) lags in the cross  correlation function of the simulated  UV -- O
light curves, in disagreement with the results of Collier et 
al. (1998). However, they also found that if the reprocessing 
was driven  by a  component with light curve similar to that 
of the UV rather than the X-ray emission (i.e. it 
contained {\sl no} high frequency variations), then the
reprocessed emission could indeed produce a CCF with UV -- O lags of 
order  of a day, in rough agreement with Collier et al. (1998), 
provided that the X-ray source was located at a height of
$\sim 3 \times 10^{14}$ cm above the disk plane. 

In the present paper we present an analysis of the campaign results of
NGC 3516 (Edelson et al. 2000) similar to that of NGC 7469 given in
BKO. However, the present analysis is more complete in several
respects: (a) The albedo of the illuminated disk is computed (rather
than assumed to be small) using the analysis of Nayakshin, Kazanas \&
Kallman (2000; hereafter NKK) which takes into account the effects of
an ionized ``skin" on the surface of the disk induced by the
action of X-rays.  (b) The effect of the presence of the black hole,
more precisely the absence of reprocessing in a region of size $R \le 3
R_S$ underneath the X-ray source is taken into account.  (c) The
calculation of the simulated light curves includes a constant
component, associated with the intrinsic emission by the geometrically
thin accretion disk, of flux consistent with observations.  The main 
effect of this component is to reduce the variability amplitude. 
(d) We compute ``standard" 
accretion disk O--UV continuum spectra for the values of the mass and
accretion rate obtained from the best fits of the model light curves
which we then compare to observations. (e) We also compute  the full 
X-ray reflection spectra which, by comparison to those observed 
by Nandra et al. (1999), yield additional constraints to the value of 
the black hole mass and the disk inclination. 
The combination of timing and spectral fits yields constraints which
are generally not commensurate with each other, thereby excluding, 
as it will be discussed below, the simplest form of this specific model.


Our paper is structured as follows: 
\S 2 contains a description of the physics involved in the 
reprocessing of X-rays by a thin disk in hydrostatic 
equilibrium and also the method for the approximate computation 
of the X-ray  albedo for different X-ray energies under these 
circumstances. Using the 
results of the albedo computation, \S 3 presents the
response functions of the {\sl reflected} X-rays
at two different energies, in order to exhibit the X-ray 
energy dependence of this effect.  
In \S 4 simulated light curves, due to reprocessing of the
X-rays observed in NGC 3516, are produced for the
wavelengths used in the monitoring campaign; 
their cross correlation functions as well 
as those with the X-rays are computed and compared to those
observed. In \S 5 we exhibit model O--UV continuum spectra
for different values of our system parameters, which we compare 
to those observed. We also compute the associated X-ray
reflection spectra, with particular emphasis in the X-ray 
reflection and \fe line features, which are compared 
to those observed. Finally in \S 6 the results are 
summarized and conclusions are drawn.

\section{Disk X-Ray Illumination and Albedo}\label{sect:skin}

A self-consistent calculation of the integrated X-ray scattering
albedo, ${\cal{A}}$, is important because the amount of the thermalized
radiation emitted in UV -- O band is a function of the X-ray albedo:
the amount of the UV -- O flux, $\fr$, due to reprocessing of the
incident X-ray flux, $\fx$, is
\begin{equation}
\fr = (1-{\cal{A}}) \fx\;.
\label{fr}
\end{equation}
It has been argued in the past that, because the typical Seyfert 1
spectra exhibit reflection continuum and \fe line features consistent
with reflection by  neutral matter, the albedo of the reflector is 
${\cal{A}} \simeq 0.1-0.2$ (e.g., Magdziarz \& Zdziarski 1995), and 
therefore it is entirely justifiable to assume this to be essentially
its correct value.

However, the recent calculations of NKK show that an ionized skin 
forms on the surface of an X-ray irradiated disk that could 
greatly influence both the albedo and the reflected spectra. 
In particular, when the skin is {\em completely}
ionized, the spectra may {\em appear} to result from reflection by
neutral-like matter because the skin itself does not leave any atomic
physics imprints on the spectra, with the Compton reflection ``hump" 
and the $\simeq 6.4$ keV Fe K$\alpha$ line then formed in the
underlying cold, neutral matter. In other words, the fact that one 
often sees neutral-like reflection and Fe lines in Seyfert 1's 
does {\em not necessarily} mean that the reflector is neutral
 and that the albedo ${\cal{A}} \ll 1$.  In fact, results of NKK 
show that the completely ionized skin can yield albedo reaching 
very high values, such that $1-{\cal{A}} \ll 1$, with the
resulting iron line still peaked at $6.4$ keV, albeit with a smaller
equivalent width (EW).  More recently, Nayakshin \& Kallman (2000)
showed that in the lamp-post geometry, the skin is almost never
completely ionized, which means that situations with $\tskin\gsim 1$
and EW $\sim $ few hundred eV (as observed in NGC~3516 by Nandra et 
al. 1999) are possible. Therefore, it does appear that an optically
thick X-ray heated skin may in fact be present in NGC~3516, and hence we need
to include it in our considerations below.

\subsection{Thomson depth of the ionized ``skin"}\label{sect:depth}

Because of the large computational overhead in the exact determination
of the integrated X-ray albedo in a time-dependent X-ray reflection
problem, we compute the albedo in an approximate way (the spectra for
static situation are calculated ``exactly'' in \S 5). The first step
in such a calculation is to determine the Thomson depth of the skin,
$\tau_s$ (for every radius). Here we follow the method of Nayakshin
(2000) with some modifications that extend the region of validity of
his results. In particular, Nayakshin (2000) assumed for simplicity
that the radiation field is roughly constant within the skin.
Strictly speaking, this assumption is appropriate only when
$\tau_s/\zeta \ll 1$, where $\zeta$ is the cosine of the incident angle
for the X-rays. Instead of this, we now use the following iterative
procedure: (1) Assume an initial value for the Thomson depth of the
skin and calculate the radiation field within the skin with the
approximate methods discussed in Sobolev (1975) and described in some
detail in \S \ref{sect:rtransfer}; (2) Find the gas pressure at the
bottom of the skin (where $P_{\rm gas} = P_c$, see Nayakshin \&
Kallman 2000); (3) Determine the geometrical location of the bottom of
the skin, $z_b$; (4) Integrate the hydrostatic balance equation from
the bottom up to obtain a new value for the ``skin" Thomson depth; (5)
Repeat the above steps until the procedure converges. This
iterative calculation is in fact similar to the iterative scheme that
one uses to solve the problem with ``exact'' numerical methods (e.g.,
NKK). However, the difference is that here we do not compute the
ionization structure of the skin (it is assumed completely ionized) and 
hence can perform the transfer of 
radiation analytically, reducing the computation time from hours
to seconds. We will report
details of this approximation in a separate future publication. These
modifications allow us to treat the optically thick situations and
also arbitrary incidence angles.

\subsection{Approximate radiation transfer}\label{sect:rtransfer}

We assume that the illuminated gas consists of two layers, of
which the top one is completely ionized while the bottom one
neutral. We can then follow the methods described in 
Sobolev (1975), developed for the treatment of radiation transfer 
in the atmospheres of planets. In this
approach, Compton scattering is assumed to be monochromatic 
(justifiable here since the photon energies we consider are much
smaller than $m_e c^2$ and the skin temperature is
only $\sim $ few keV). The cold, neutral medium of the bottom layer
is ascribed an albedo ${\cal{A}}_c$, appropriate to neutral matter 
(${\cal{A}}_c=0.2$). Following the derivation given in Chapter 8 of 
Sobolev (1975), especially his \S 8.6, one can easily derive an 
approximate expression for the distribution of the radiation field 
at any depth within a skin of a given Thomson depth $\tskin$. 
This local radiation field determines the Compton temperature
and the angle-integrated intensity of the radiation, which are needed
in steps (2) and (4) of the iterative procedure of the calculation 
of $\tskin$ described above.

The same procedure can be also used for the approximate calculation of
the energy dependence of the reflected X-ray intensity 
by using an albedo for the bottom, neutral layer 
which is energy dependent. Because we know from our earlier, exact 
calculations (see NKK), that the temperature of this layer 
is always sufficiently low that the gas is weakly ionized, one can 
simply use the albedo of the neutral matter, given, for example, 
by Magdziarz \& Zdziarski (1995). 
The reflected X-ray flux at an energy $E$ is then given 
by equation 8.100 of Sobolev (1975, with his $x_1 = 0$ because we 
assume that the scattering is isotropic), with the albedo ${\cal{A}} = 
{\cal{A}}_c(E)$ as given by Magdziarz \& Zdziarski (1995), and the 
notation changed appropriately to ours (e.g., $\tau_0\equiv \tskin$, 
etc.). Clearly, this approach is not accurate for sharp spectral 
features, such as the \fe line, or for high photon energies, i.e., 
$E\gsim 50$ keV or so, but it is nonetheless adequate for our purposes
(see \S \ref{sect:xsponce}).

\subsection{The Albedo}\label{sect:albedo}

Using the radiative transfer approach described in \S
\ref{sect:rtransfer}, it is easy to show that the albedo due to the
Compton scatterings in the skin is given by
\begin{equation}
{\cal{A}}_{s0} = 1 - {2 +3\zeta\over 4 +
3\tskin}\, - \, {2 - 3\zeta\over 4 + 3\tskin}\; e^{-\tskin/\zeta}\;.
\label{albs}
\end{equation}
where $\zeta$ is the cosine of the incidence angle, while 
$\tskin$ the total Thomson depth
of the skin. This expression assumes that all the X-rays incident 
on the cold material below the ionized skin are absorbed. 
In reality, 10-20 \% is 
reflected (see MZ95), and thus we approximate this situation by
writing
\begin{equation}
{\cal{A}}_s = {{\cal{A}}_{s0} + {\cal{A}}_c \over 1+ {\cal{A}}_c}\;,
\label{albt}
\end{equation}
where ${\cal{A}}_c$ is the integrated albedo of the cold material, which we
assume to have the value ${\cal{A}}_c = 0.2$. The above approximate 
expression has the correct behavior for both optically thin (i.e., 
$\tau_s\ll 1$, when ${\cal{A}}_s\simeq {\cal{A}}_{s0} + {\cal{A}}_c$) 
and thick limits (${\cal{A}}_s\simeq {\cal{A}}_{s0}$) and
will suffice for our present study.

\section{The X-Ray Reflection Response}\label{sect:xsponce}

With an approximate expression for the reflected X-ray intensity 
at hand we can now discuss
the response of the X-rays reflected (rather than reprocessed) by the 
disk as a function of photon energy. To simplify the 
treatment we will consider the response at only two photon energies,
namely $E = 1$ and $20$ keV. The lower value is representative of the 
behavior of soft X-rays that are largely absorbed by the neutral 
reflector (but may be reflected if $\tskin\gsim 1$), whereas the higher
one is representative of the behavior of  hard photons that are mostly 
reflected rather than absorbed.  It is of interest to examine whether 
such a different behavior would result in effects between these two
bands which could be observed in the time domain. Even  though no 
such analysis has been done for  the multiwavelength campaign data
of NGC~3516 that we are discussing here, we point out that these
timing correlations within the X-ray band itself are an important
property of the model and future observations should be planned as to
extract these correlations in addition to those of the O--UV band with
the X-rays.

The geometry of the source considered herein is identical to that
discussed in BKO: A point-like X-ray source located at a height 
$h_X$ above the accretion disk and in particular above the compact 
object (i.e. the lamp-post model). To present a more precise treatment
than BKO we also consider that the reprocessing surface
does not extend to the foot of the vertical from the X-ray source 
onto the disk plane, but only to three \sw radii, $R_S$.  The presence 
of the black hole, or rather the absence of the reprocessing disk 
at radii $r \le 3 \, R_S$ is of some significance for the response 
function, especially when $h_X \lsim 3 \,R_S$; under these conditions
both the minimum lag associated with X-ray reprocessing off the 
accretion disk and the fraction of the disk reprocessed X-ray luminosity 
are significantly different from the case that no such hole is 
present. Finally, as in BKO, the observer is considered to be at 
a position of latitude $\theta$ above the disk. 

As shown in BKO the response function of an infinite plane to 
an X-ray source at a height $h_X$ in the direction of the observer
(at latitude $\theta$), as a function of the time lag $\tau$ and
the radial distance $R$ from the black hole, is
\begin{equation}
A(R, \tau) = \cases{ 2 \, R/\sqrt{(\tau - \tau_l)(\tau_t -\tau)}
 & if $\tau_l < \tau < \tau_t$ \cr
0 & otherwise \cr}~.
\end{equation}
where $\tau_l = D_l/c$ and $\tau_t = D_t/c$ are the leading and trailing
lags respectively at which the ellipses of constant lag (the intersection 
of paraboloids of constant lag with the disk plane) intersect the 
circle of a given constant radius (constant X-ray flux). The 
quantities $D_l$ and $D_t$ are given by  the expressions
\begin{eqnarray}
D_l &=& \sqrt{R^2 + h_X^2} - R \, cos \, \theta + h_X \, sin \, \theta ~~,\\
D_t &=& \sqrt{R^2 + h_X^2} + R \, cos \, \theta + h_X \, 
sin \, \theta
\end{eqnarray}

The reflected X-ray flux, $F_{XR}$, at a given radius $R$ 
depends on the local angle specific albedo ${\cal{A}}(E, \theta)$, 
which, as discussed above,
is a function of both the X-ray energy $E$ and the observer 
latitude angle $\theta$. The expression for $F_{XR}$ is
\begin{equation}
F_{XR}(E,t) = \frac{L_X(t) \, {\cal{A}}(E)}{4\, \pi \,(h_X^2 + R^2)} 
\frac{h_X}{(h_X^2 + R^2)^{1/2}} 
\end{equation}
where $L_X(t)$ is the luminosity of the X-ray source, which is a function
of time. Then, the reflected X-ray luminosity as a function of time, 
$L_{XR}(E,t)$, is obtained by integrating the reflected X-ray 
flux over all lags $\tau$ and over the entire area of the disk. 
Due to the energy dependence of the albedo the above quantity depends 
on the energy of the X-ray photons. Thus,

\begin{equation}
L_{XR}(E,t) = \int_0^{\infty} d R \, \int_{\tau_l}^{\tau_t} d \tau \, 
A(R, \tau) \, F_{XR}(E, t - \tau)
\end{equation}

The presence of a minimum disk radius at $R=3\, R_S$ affects the
lower limit of the lower limit, $\tau_l$, of the time integration in 
the above equation, which is given by the maximum value 
of $D_l(R)/c$ and $D_l(R = 3R_S)/c$. The effects of the absence of the
disk at $R \le 3R_S$ are then taken into account by considering 
the proper lower limit and that the integral is zero when $\tau_l 
\ge \tau_t$. As in BKO the time and lags are measured here in 
units of $h_X/c$.

In order  to assess the effects of the different parameters 
associated with the calculation of the albedo and the disk -- 
X-ray source geometry on the X-ray reflection  by the disk, 
we present its response, i.e. the flux as a function of time 
resulting from a narrow -- in time -- pulse of X-rays for a 
variety of combinations of these parameters. Because
the albedo is in addition a function of the X-ray photon energy, 
we also consider in each case the disk response at two different 
photon energies, namely $E = 1, ~20$ keV.

\begin{figure*}[ht]
\centerline{\psfig{file=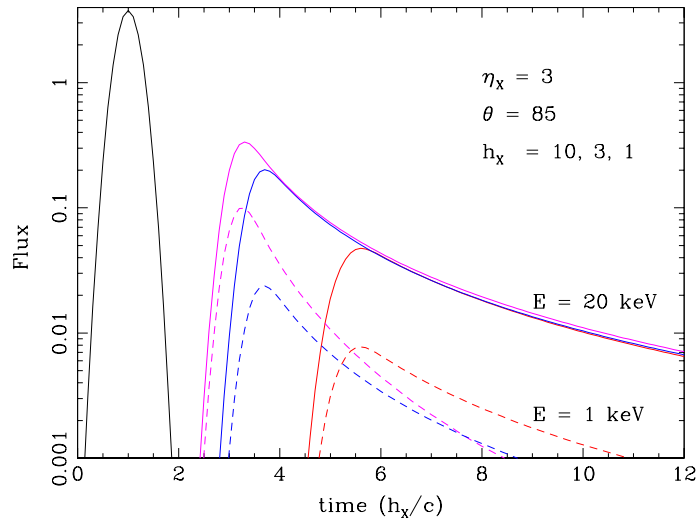,width=.55\textwidth,angle=-90}}
\caption{The response of the disk to a Gaussian X-ray impulse 
(the curve having a maximum at $t = 1$) as a function 
of the source height $h_X$ (in \sw radii) from the disk plane for
two different photon energies $E = 1$ keV (dashed curves)  and 
$E= 20$ keV (solid curves). The disk is almost ``face-on" 
($\theta = 85$ degrees) with $\eta_X = L_X/L_d =3$.}
\label{fig:hx}
\end{figure*}

The form of the X-ray pulse assumed is that of a Gaussian of unit
area, i.e. $L_X(t) = exp[-(t - t_0)^2/\tau_1^2]/ \sqrt{2 \pi
\tau_1^2}$. The pulse width $\tau_1$ is a free parameter taking (in
units of $h_X/c$) the values $\tau_1 = 0.3$. The other parameters
used are the ratio of the X-ray to disk luminosities, $\eta_X$, the
latitude angle of the observer, $\theta$, and the height, $h_X$, of
the X-ray source above the disk plane (in \sw radii). While the mass
$M$ of the black hole does not figure in the geometry of X-ray
scattering it is implicitly involved in the computation of the albedo,
because it determines the absolute value of the X-ray flux on the disk
surface, which in turn determines its temperature.  In all the runs
presented in this section the black hole mass was held constant at the
value $M = 10^8$ M$_{\odot}$, while the value of $\dot M$ was kept at
$\dot M = 0.003$ in units of the Eddington value, yielding a
luminosity in general agreement with that of NGC 3516.

Figure \ref{fig:hx} presents the response function of a nearly 
face-on disk ($\theta = 85$ degrees) of $\eta_X=3$ for three different 
values of the X-ray source height $h_X = 10, 3, 1$ (in descending 
order in the figure) and for two different X-ray photon energies 
$E = 20$ keV (solid curves) and $E= 1$ keV (dashed curves), along
with the input impulse of width $\tau_1 = 0.3$. The 
figure exhibits clearly the strong dependence of the albedo
on the photon energy. Also apparent is the effect of decreasing
the X-ray source height $h_X$ on both the response amplitude
and lag at which the response peaks. This effect is due to 
the absence of reprocessing matter at $R < 3 \, R_S$. Because
the disk albedo at photon energy $E = 20$ keV is almost 
independent of the incidence -- reflection geometry and 
structure of the ionized skin on the disk surface (see upper
panel of fig. 7 of NKK), the
response curves overlap at large values of the time (lag).
This is clearly not the case with the lower energy
photons ($E = 1$ keV), for which the shape of the 
response changes significantly with $h_X$, because of the 
dependence of the value of $\tskin$ on this parameter.

The difference in response between the X-ray photons
of different energies suggests that the presence of 
the cold disk in the vicinity of the X-ray source would 
modify the spectrum of the observed radiation, in the
sense that the ratio of soft to hard photons received
by the observer at infinity will be smaller than that 
produced by the source, due to the larger fraction 
of soft photons absorbed by the disk. This produces the well 
known ``reflection" component (Basko et al. 1974; 
Lightman \& White 1988; Magdziarz \& Zdziarski 1995), which 
has apparently been 
observed in AGN (Nandra \& Pounds 1994). However, in
addition to the ``hardening" of the (time integrated) 
spectrum due to the reflected component, the time delay information 
is also preserved in the signal: because a predominantly 
larger fraction of the hard relative to the soft photons reach the
observer after reflection, i.e. after time $\sim h_X/c$,
this geometric arrangement should lead to a broader 
autocorrelation function of the response with photon energy
and therefore to lags in the cross-spectrum of the 
source between soft and hard energy bands, even though
the CCF of the same bands peaks at zero lag.

It is of interest to remind the reader that hard lags 
in the cross-spectrum have 
indeed been observed in galactic X-ray sources,
most notably in Cyg X-1 (Miyamoto et al. 1988; Nowak et 
al. 1998). These have been interpreted
as due to Compton scattering in an extended hot 
corona (Kazanas, Hua \& Titarchuk 1997; Hua, Kazanas 
\& Cui 1999), or systematic hardening of the X-ray 
emission with time for a variety of reasons (Poutanen \& 
Fabian 1999; B\"ottcher \& Liang 1999). The difference in X-ray 
energy albedo in the geometry discussed above represents 
another process which would also induce lags of the correct 
sign in the X-ray light curves. However, while the lag sign 
is correct, their magnitude is much smaller than 
those observed. This could be remedied by either by removing 
the X-ray source sufficiently far away from the disk or by 
considering the photons which scatter at larger disk radii. 
The first option requires the {\sl ad hoc} production of 
X-rays at very large distances above the disk; while one 
could contrive scenaria that would allow this, we refrain 
from doing so here. The second option produces only a very 
small number of photons which would exhibit these lags, since 
the X-ray flux reflected by the large radii portions of the disk
decreases like $R^{-2}$; thus, barring a significant change
of the disk geometry with distance, it would be hard to discern 
the delays between these photons among those  which arrive 
to the observer directly from the source.

\begin{figure*}[ht]
\centerline{\psfig{file=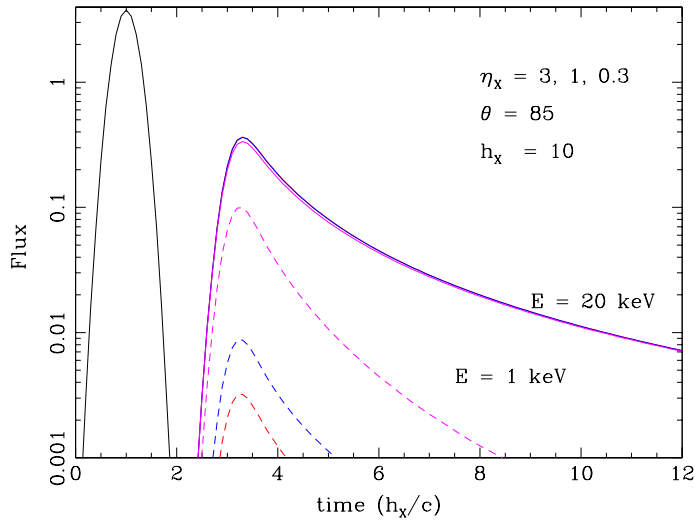,width=.55\textwidth,angle=-90}}
\caption{The response of the disk to a Gaussian X-ray impulse 
(the curve having a maximum at $t = 1$) as a function 
of the X-ray to disk luminosities $\eta_X = L_X/L_d$ for a 
given source height $h_X = 10$ and observer latitude 
($\theta = 85$ degrees) for
two different photon energies $E = 1$ keV (dashed curves)  and 
$E= 20$ keV (solid curves). Two of the $E = 20$ keV curves overlap
completely, while the amplitude of the $E = 1$ keV curves decreases
with decreasing $\eta_X$.}
\label{fig:etax}
\end{figure*}

Figure \ref{fig:etax} exhibits the effects of changing the 
ratio $\eta_X = L_X/L_d$ of the X-ray, $L_X$, to the disk, $L_d$,
luminosities. These runs were produced assuming a constant 
value for the X-ray source height $h_X = 10$ above the disk 
and an  observer latitude $\theta = 85$ degrees. As in
Figure \ref{fig:hx}, the dashed and solid curves correspond
to photon energies 1 and 20 keV respectively. As in 
Figure \ref{fig:hx}, here too the response of the 20 
keV photons is independent of 
the value of $\eta_X$, indicating that the albedo at this energy
is independent of details of the structure of the ionized 
skin. However, this is not the case with the photons of 
$E = 1$ keV, as it is clearly evident in the figure. For these
photons, the decrease in albedo with decreasing $\eta_X$ leads
to a corresponding substantial reduction in the reflected 
flux at this energy. The response curves peak at a lag 
$\tau \simeq 2 h_X/c$ from the injection of the X-ray pulse,
as expected in reflection by an infinite plane for an observer
right above the source, indicating that for the chosen value
of $\eta_X$ the effects of the absence of the disk for 
$r \le 3 R_S$ are indeed small. 

\begin{figure*}[th]
\centerline{\psfig{file=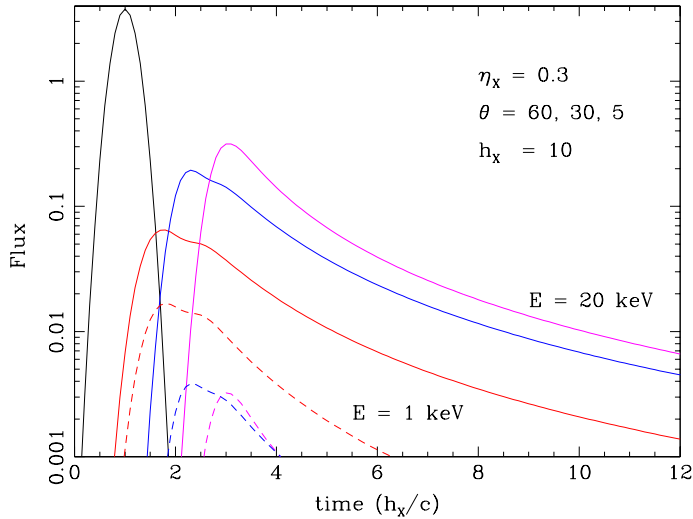,width=.55\textwidth,angle=-90}}
\caption{The response of the disk to a Gaussian X-ray impulse 
(the curve having a maximum at $t = 1$) as a function 
of the observer latitude angle $\theta$, for $\theta =
60, 30, 5$ degrees and for two different photon energies 
$E = 1$ keV (dashed curves)  and $E= 20$ keV (solid curves).
The values of the other parameters were $\eta_X = 0.3$ and
$h_X = 10$. }
\label{fig:angle}
\end{figure*}

The effects of changing the observer latitude $\theta$ on the
response function are shown in Figure \ref{fig:angle}.
In computing these curves, the values $\eta_X = 0.3$ and 
$h_X = 10$ were used, with the usual notation for solid and
dashed curves. Decreasing the observer latitude impacts
differently the amplitude of the response for the soft (1 keV) 
and hard (20 keV)  X-ray photons: The response of the latter
decreases because of the decrease of the disk area projected
along the observer's line of sight. On the other hand,
the response of the soft photons does increase with decreasing
$\theta$ because of the accompanying increase in albedo 
at these energies which more than compensates for the 
decrease in projected area. One can even discern, for the 
smallest value of the angle, a secondary peak at lag $\tau 
\simeq 2.5 $, roughly $h_X/c$ away from the primary one, 
induced by the
change of albedo with the scattering photon geometry.
Apparent in  the figure is also the expected shift of the peak 
of the response function to smaller lags with increasing 
inclination angle (decreasing $\theta$), due to the fact that 
the reflected X-rays can reach the observer with much reduced 
lag as $\theta$ in this case. As $\theta \rightarrow 0$ the 
lag in the X-ray response peak should also decrease to zero. 

\section{The X-ray Reprocessing}\label{sect:ouv}

The fraction $1 - \cal{A}$ of the X-ray radiation which is not 
reflected by the disk is absorbed and reprocessed, presumably
into thermal radiation, leading to the expected (and 
presumably seen) fluctuations at the continuum emission at
UV -- O wavelengths. The expected fluctuations can be 
easily computed by computing the temperature variations 
on the surface of the disk induced by the variable X-ray
flux as a function of time (lag) and radius, and integrating
the flux at a given wavelength over the entire disk area,
in a fashion similar to that used in the previous section
to compute the X-ray reflection as a function of time.

\begin{figure*}[ht]
\centerline{\psfig{file=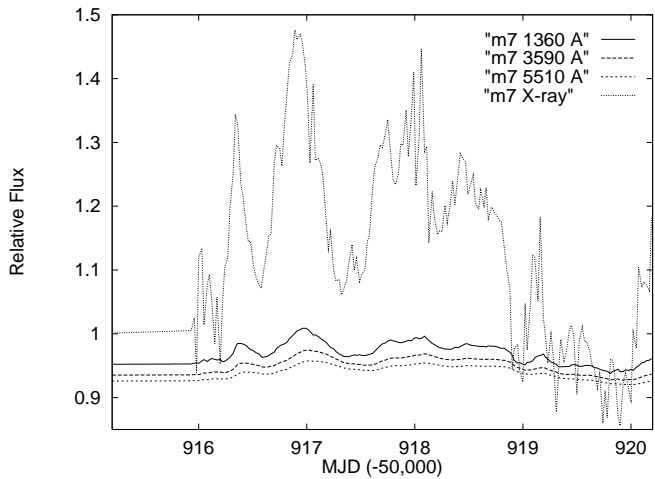,width=.55\textwidth}}
\caption{The X-ray light curve (dotted line) along with the model
light curves at \l 1360 \AA~ (solid line), \l 3590 \AA~ (long-dashed
line) and \l 5510 \AA~ (short-dashed line) for $M = 10^7$ M$_{\odot}$
$\theta = 30^{\circ}$, $\eta_X = 0.3$. }
\label{fig:m7xrelfl}
\end{figure*}

\begin{figure*}[hbt]
\centerline{\psfig{file=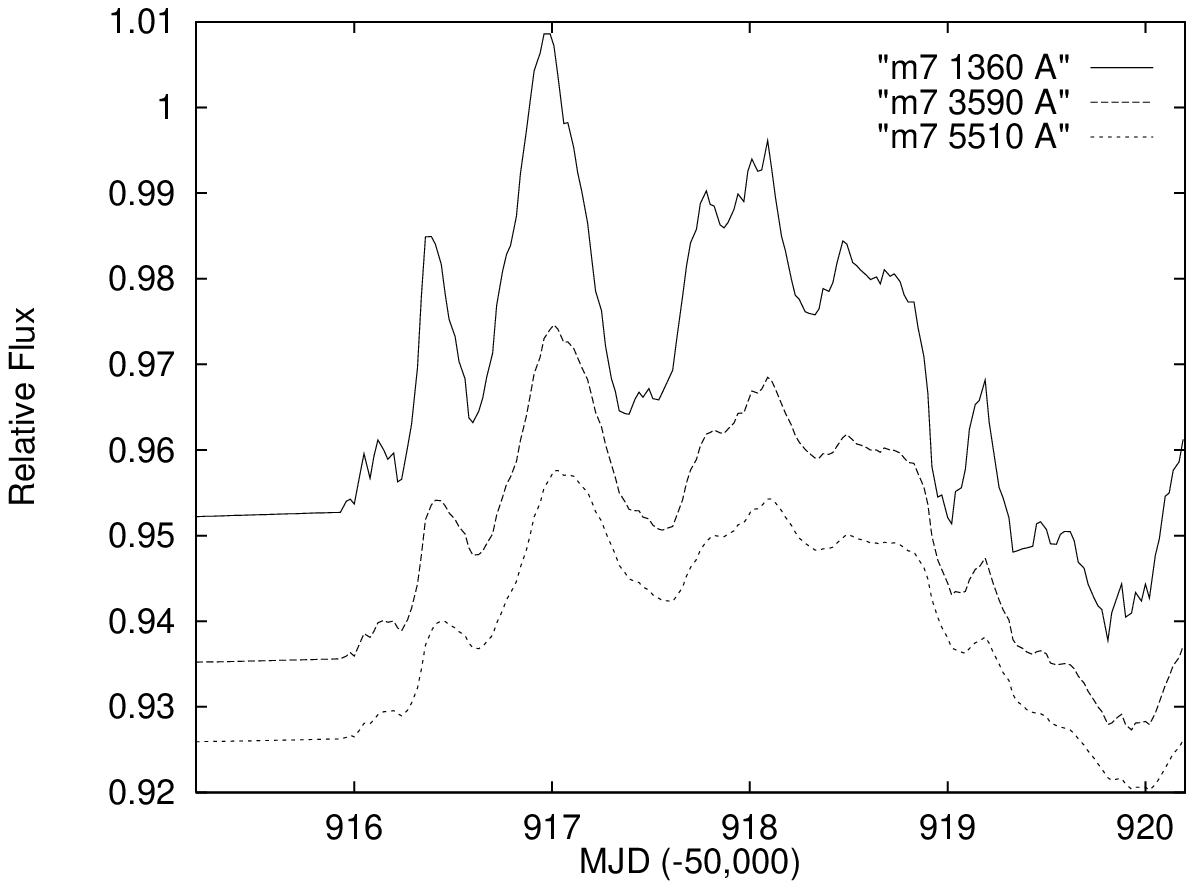,width=.55\textwidth}}
\caption{Expanded version of Fig. (\ref{fig:m7xrelfl}) showing only
the model light curves of the reprocessed radiation. The wavelength
assignments are the same. }
\label{fig:m7relfl}
\end{figure*}

The temperature of the disk at a given distance $R$ from the compact 
object is determined by the sum of the fluxes from: (a) The 
reprocessed  X-rays and (b) The intrinsic flux of the geometrically thin
disk. We assume the ratio of the X-ray to the disk luminosity 
to be $\eta_X = L_X/L_d$, while the total luminosity $L = L_X + L_d$.
Then we have the following relations between $L, ~L_X,~L_d$
\begin{equation}
 L_X = \frac{\eta_X}{1 +  \eta_X} L \;, ~~~~~~ L_d = \frac{1}{1 +  \eta_X} L
\end{equation}

Viscous dissipation provides for accretion onto
a \sw black hole  a total local radiant 
flux $F$ of magnitude (e.g. Shapiro \& Teukolsky 1983)
\begin{equation}
F = \frac{3 G M \dot M}{8 \pi R^3} \left[1 - \left(\frac{R_i}{R}
\right)^{1/2} \right]
\end{equation}
where $M$ is the mass of  the accreting black hole, $\dot M$ is the 
accretion rate and $R_i$ the inner edge of the accretion disk,
with $R_i = 3 \, R_S $ in the case of a \sw black hole considered
here. Let $\epsilon$ denote the efficiency of the thin disk
in converting mass flux to radiation; then the local radiant 
flux due to the thin disk thermal emission, $F_d$, and the 
corresponding temperature $T_d$ (assuming black body emission) are
given by the relation 
\begin{equation}
F_d = \sigma T_d^4 = \frac{3}{16 \pi \, R_S^2}\frac{L_d}{\epsilon}
\left[1 - \left(\frac{x_i}{x}\right)^{1/2} \right]x^{-3}
\end{equation}
where $R_S = 3 \times 10^{13} \, M_8$ cm is the \sw radius of the
black hole ($M_8$ is the black hole mass measured in units 
of $10^8$ M$_{\odot}$)
and $x$ is the radius normalized to $R_S$, with $x_i = 3$ and
$\epsilon = 0.06$ in the case of a disk around a \sw black hole.
On the other hand, the reprocessed X-ray flux at radius $R$ on 
the disk and the temperature of the associated thermal emission are 
given by 
\begin{equation}
F_X = \sigma \, T_X^4 = \frac{L_X \, (1 - \cal{A})}{4\, \pi \,(h_X^2 + R^2)} 
\frac{h_X}{(h_X^2 + R^2)^{1/2}} 
\end{equation}
where $L_X$ is the luminosity of the X-ray source and $\cal{A}$ is
the energy integrated, angle averaged albedo of the disk. 

In applying the above considerations to models of the NGC~3516 
campaign, because the X-ray luminosity is variable 
while that of the disk $L_d$ is assumed to be constant, the value 
of their ratio $\eta_X (=0.3)$ referred to in the rest of the 
paper is that observed 
at the beginning of the monitoring campaign. 
The instantaneous value of the temperature used in the computation
of the variable flux is then given by
\begin{equation}
T(R,t) = \left[ \frac{F_X(R,t)}{\sigma} + \frac{F_d(R)}{\sigma}\right]^{1/4}
\end{equation}
With the time dependent temperature as a function of radius 
at hand, the flux at a given wavelength \ls is obtained by 
first integrating this emission at a given radius $R$ over 
all lags $\tau$ and then over all disk radii, to get 
\begin{equation}
f_{\l}(t) = \int_0^{\infty} d R \, \int_{\tau_l}^{\tau_t} 
d \tau \, A(R, \tau) \, B_{\l} [T(R, t - \tau)] \, 
\label{ouvlcv}
\end{equation}
where $B_{\l} (T)$ is the black body emissivity of temperature $T$
at wavelength \l. 

\begin{figure*}[ht]
\centerline{\psfig{file=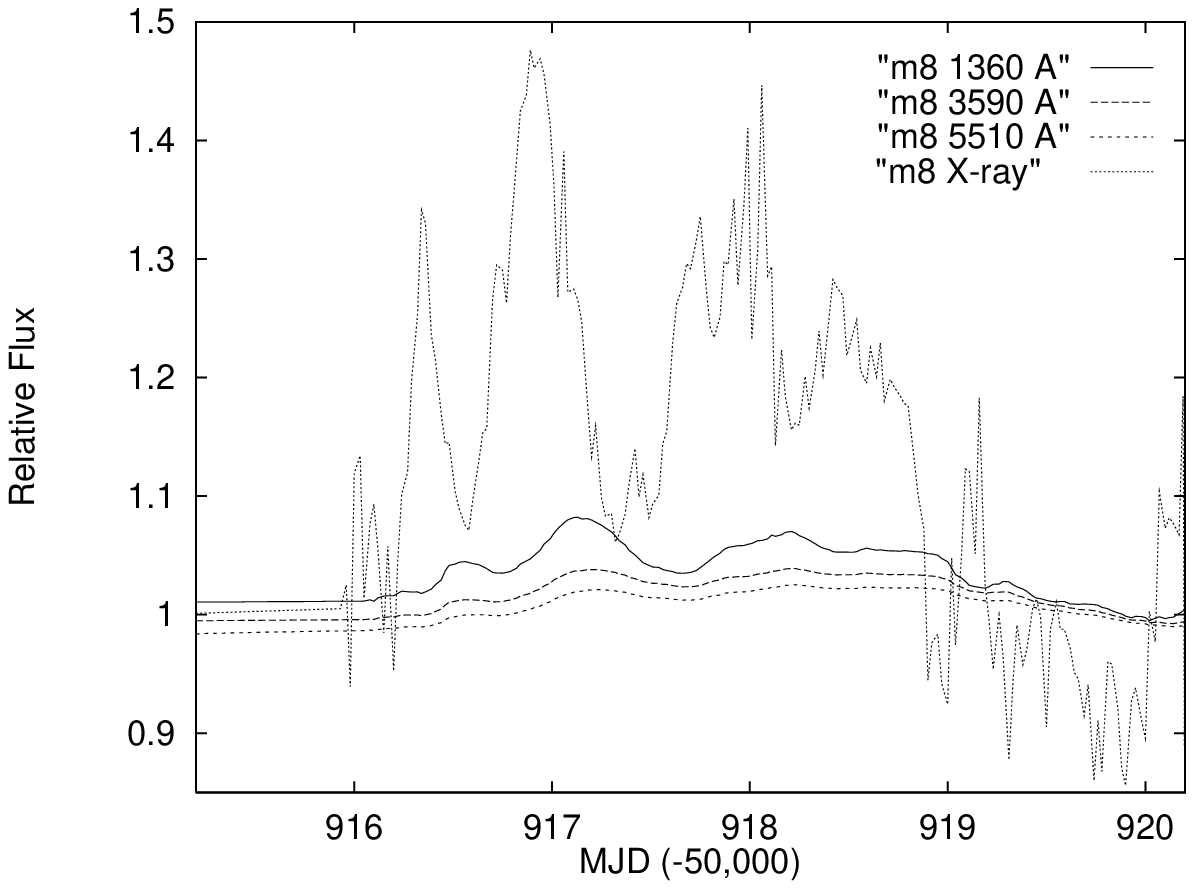,width=.55\textwidth}}
\caption{The X-ray light curve (dotted line) along with the model
light curves at \l 1360 \AA~ (solid line), \l 3590 \AA~ (long-dashed
line) and \l 5510 \AA~ (short-dashed line) for $M = 10^8$ M$_{\odot}$
$\theta = 30^{\circ}$, $\eta_X = 0.3$. }
\label{fig:m8xrelfl}
\end{figure*}

\begin{figure*}[hbt]
\centerline{\psfig{file=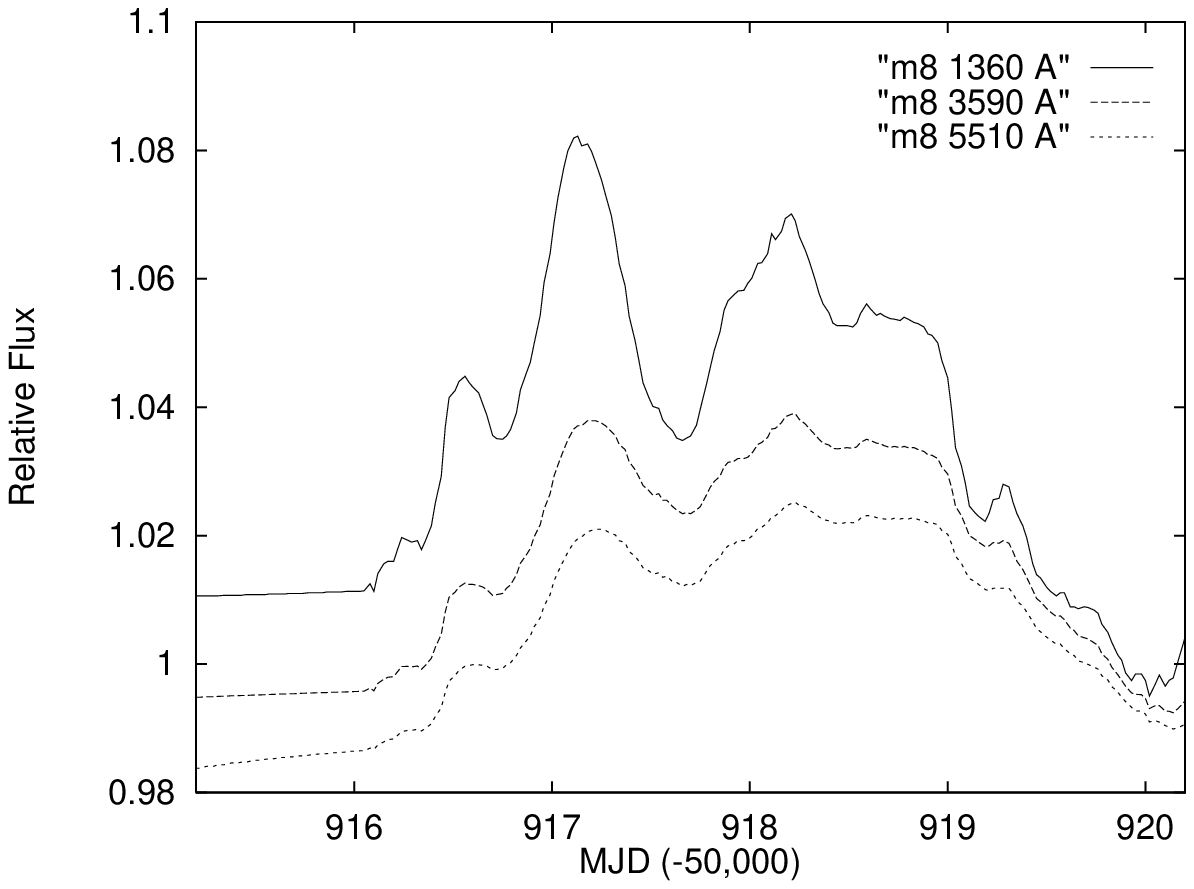,width=.55\textwidth}}
\caption{Expanded version of Fig. (\ref{fig:m8xrelfl}) showing only
the model light curves of the reprocessed radiation. The wavelength
assignments are the same. }
\label{fig:m8relfl}
\end{figure*}

Using the above expression (Eq. \ref{ouvlcv}) and the X-ray 
light curve of NGC 3516 of Edelson at al. (2000), we produced 
model light curves at a number of wavelengths in the UV and 
optical part of the spectrum, specifically at \l\ls 1360, 3590 
and 5510 \AA~ to allow a direct comparison with those observed. 
The X-ray light curve used in Eq. (\ref{ouvlcv}) was obtained 
by a direct logarithmic interpolation of the light curve of 
Edelson et al. (2000), graciously provided to us by Dr. K. Nandra.
One should note that the use of Eq. (\ref{ouvlcv}) in 
computing the model light curves requires the knowledge of 
the X-ray flux at times prior to that of day 915.8, at which 
the monitoring campaign began. Given  that no information
about this flux is available, we assumed it to be constant
at the level it had at the beginning of the campaign. In 
addition to this problem, one must also allow this constant 
flux persist sufficiently long for the reprocessed radiation to 
reach a steady state level. Not taking this into consideration 
may result in a model light curve dominated by the X-ray 
``turn-on" phase. We have experimented with 
the ``turn-on" time and found that for the longest wavelength
(\l 5590 \AA) and for the largest value of the mass ($M_8 =1$), 
we should turn the source on at day 912 in order to achieve 
a steady emission at this wavelength by day 915.8.

In computing the model light curves we assumed that the X-ray spectrum
of NGC 3516 is a power law, i.e. $F_X \propto E^{-\alpha}$ erg \s
keV$^{-1}$, $\alpha \simeq 0.5$, (Nandra et al. 1999), which extends
to energies $E \simeq 100$ keV, thereby yielding for $\eta_X = L_X/L_d
= 0.3$.  Nandra et al. (1999), who presented the spectral analysis of
the X-ray observations of this campaign, concluded on the basis of the
\fe line profiles that the disk inclination is small ($\theta$ is
large), so that the value of $\eta_X (= 0.3)$ inferred above is a fair
representation of the true value of this parameter because no
significant absorption in an obscuring torus is expected for such
small inclination angles. We therefore use this value for $\eta_X$ in
the remainder of our paper, irrespective of the value of the latitude
angle, $\theta$, used.

With the value of $\eta_X$ fixed we study the  effects of the 
remaining two parameters $\theta, ~M$ on the model light curves
due to X-ray reprocessing. Figure \ref{fig:m7xrelfl} exhibits 
the relative amplitude of 
the observed X-ray light curve along with those of the 
model light curves due to X-ray reprocessing for 
the three wavelengths considered, namely \l\ls 1360, 3590  
and 5510 \AA~ for a black hole mass $M = 10^7$ M$_{\odot}$,
$\eta_X=0.3$ and $\theta = 30^{\circ}$. Figure \ref{fig:m7relfl} 
is an expanded version of Figure \ref{fig:m7xrelfl}, exhibiting 
only the relative amplitudes of the model light curves. Similarly, 
Figures \ref{fig:m8xrelfl} and \ref{fig:m8relfl} exhibit the 
relative amplitudes of the same light curves for the same values
of $\eta_X$ and $\theta$ but a black hole mass $M = 10^8$ M$_{\odot}$.

\begin{figure*}[hbt]
\centerline{\psfig{file=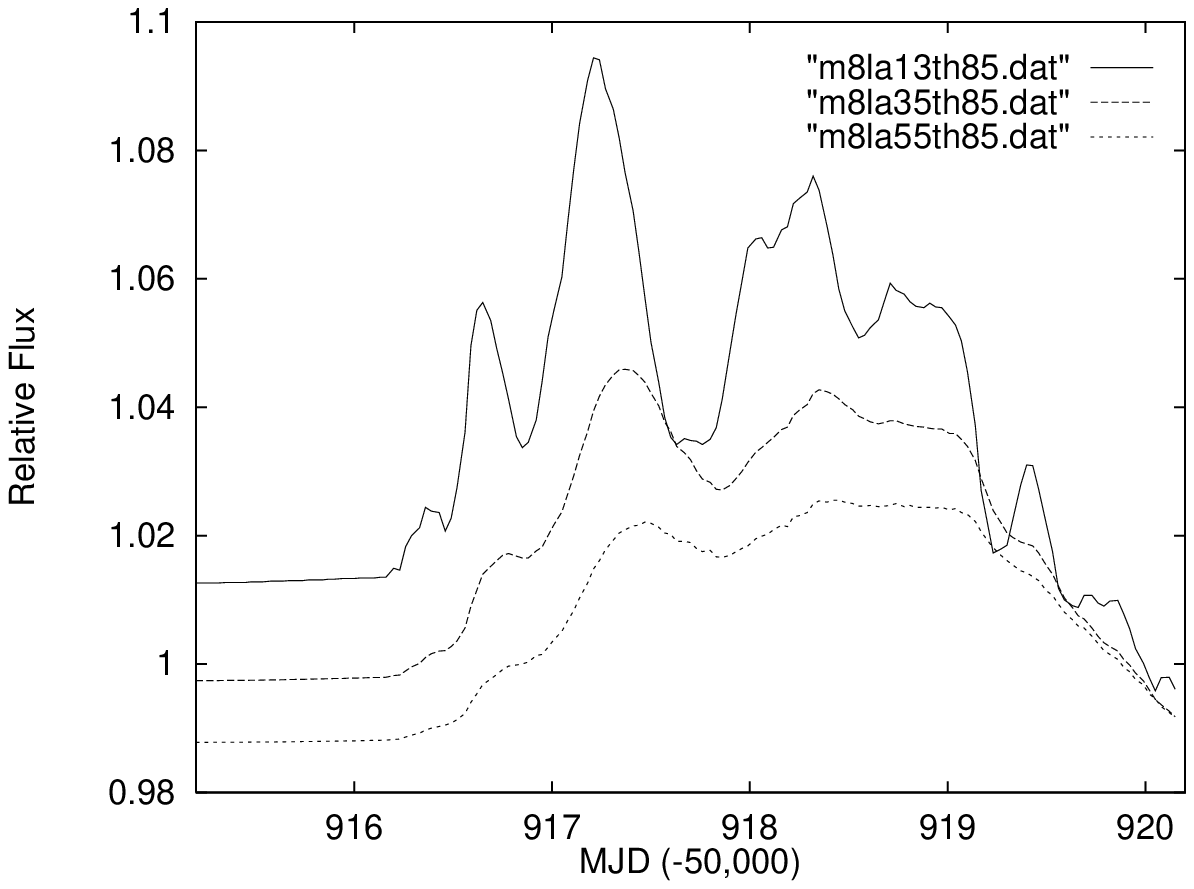,width=.55\textwidth}}
\caption{Same as Fig. (\ref{fig:m8relfl}) for $\theta = 85^{\circ}$. }
\label{fig:m8th85}
\end{figure*}

\begin{figure*}[hbt]
\centerline{\psfig{file=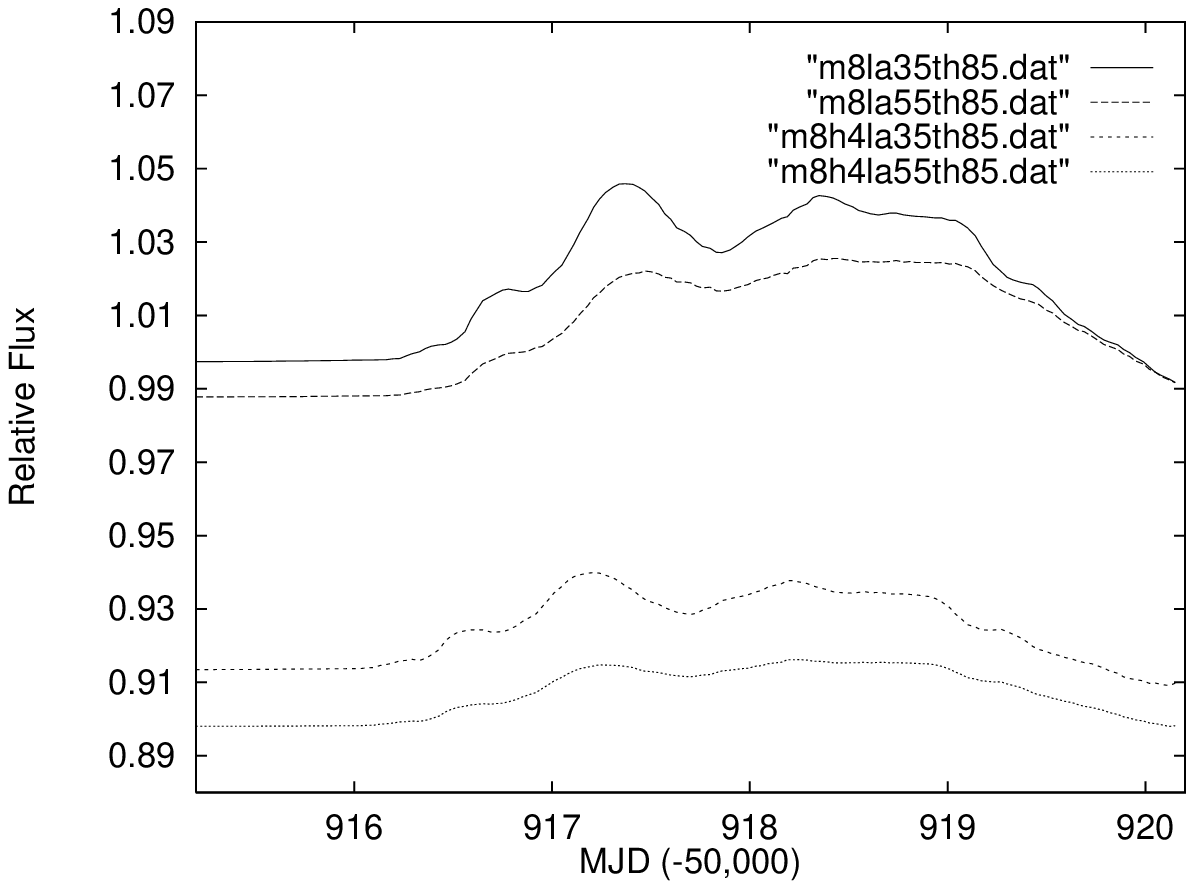,width=.55\textwidth}}
\caption{Two sets of model light curves exhibiting the effects of the 
height of the X-ray source $h_X$. The top two curves correspond to 
\l\ls 3590, 5510 \AA, for $M = 10^8$ M$_{\odot}$, $\theta = 85^{\circ}$ 
and $h_X =10$. The bottom curves correspond $h_X =4$ and identical 
values for the rest of the parameters. }
\label{fig:m8hxth}
\end{figure*}

We also explored the effects of the observer latitude angle $\theta$
and the height of the source $h_X$ on the resulting light curves. 
Figure \ref{fig:m8th85} has identical parameters to those of 
Figure \ref{fig:m8relfl} except for the observer latitude which is 
$\theta = 85^{\circ}$. There is an apparent increase in the lag of 
the reprocessed light curves relative to those of $\theta = 
30^{\circ}$, as well as a small increase in the variability 
amplitude, which for \ls 3590 \AA~ reaches $\simeq 5\%$. Finally,
for reasons which will be explained in the next section (the need
for higher X-ray flux at small $R$ to improve the spectral fits) 
we have also produced
light curves for the same almost face-on models ($\theta = 85^{\circ}$)
but smaller distance of the X-ray source from the disk ($h_X = 4$).
The light curve for this set of parameters are shown (along with 
those of $h_X = 10$ for comparison) in Figure \ref{fig:m8hxth}.
The effects of reducing $h_X$ are a small decrease in the respective
lags and a substantial decrease in the level of variability to 
the $\simeq 2$ \% level.

The most apparent feature of the model light curves is the
smallness of their variability amplitudes, given the large 
fluctuations ($\sim 70\%)$ of the X-ray flux. 
This is the result of the combination of ``spreading" the 
reprocessed emission over a large range of radii 
and in addition of ``diluting" its variations by the (presumably)
constant flux intrinsic to the accretion disk. As shown in figures 
\ref{fig:m7relfl}, \ref{fig:m8relfl} the variation amplitudes 
for \l\ls 3590 and 5510 \AA are of order $\simeq 3 - 5
\%$; while they are larger and more 
easily discernible at \l 1360 \AA~ the observations in this 
wavelength are not reliable due to the influence of SAA on the
UV detectors.  Therefore, one cannot exclude the X-ray 
reprocessing model on the basis of the observed optical 
variability amplitudes. 

However, despite the similarity in variability amplitudes
between the observed and the model light curves, 
closer inspection reveals several differences 
between them that put the entire scheme into question. 
This has to do with the absence of O -- UV response associated with 
the flare in X-ray flux which peaks near day 917. While this feature 
is evident and persists in all model light curves presented in the 
above figures, it appears to be absent in the observed light
curves (see  fig. 2 of Edelson et Al. 2000). The variability in
the observed light curves seems to be completely dominated 
by the much broader and of lower peak-flux feature between days 
917.3 and 920. One should note that the model light curves exhibit 
an increasing preponderance of this lesser flux, broader feature 
with the wavelength of reprocessed emission and black hole mass. 
Such a  behavior in the time profiles of 
reprocessed emission, in response to low-broad {\it vs.}  
sharp-narrow  features in the light curve of the driving flux,
indicates that the size of the reprocessing region is larger 
than the duration of the ``narrow" feature and more in line 
with that of ``broad" one. This trend is apparent in figures 
\ref{fig:m7relfl} and \ref{fig:m8relfl} in which the amplitude 
of the feature with peak flux near day 917 becomes less pronounced 
relative to that extending from day 917.3 to 920, even though its peak 
flux (at day 918) is smaller. This suggests that using an even larger
value for the black hole mass (actually, a value  $M = 3 \times 
10^8$  M$_{\odot}$ for the black hole mass would
produce a better fit to the UV spectrum -- see next section) 
might reproduce the observed amplitude ratio of these two 
features; however, such an increase in the mass would further increase the 
already discernible lags between the X-ray and O -- UV emission
(see figure \ref{fig:m8relfl}) which are not present in the data 
(see figure 2 of Edelson et al. 2000). The above discussion (and associated
figures), in essence, casts  the arguments against reprocessing made
by Edelson et al. (2000) in a quantitative form. 

\begin{figure*}[hbt]
\centerline{\psfig{file=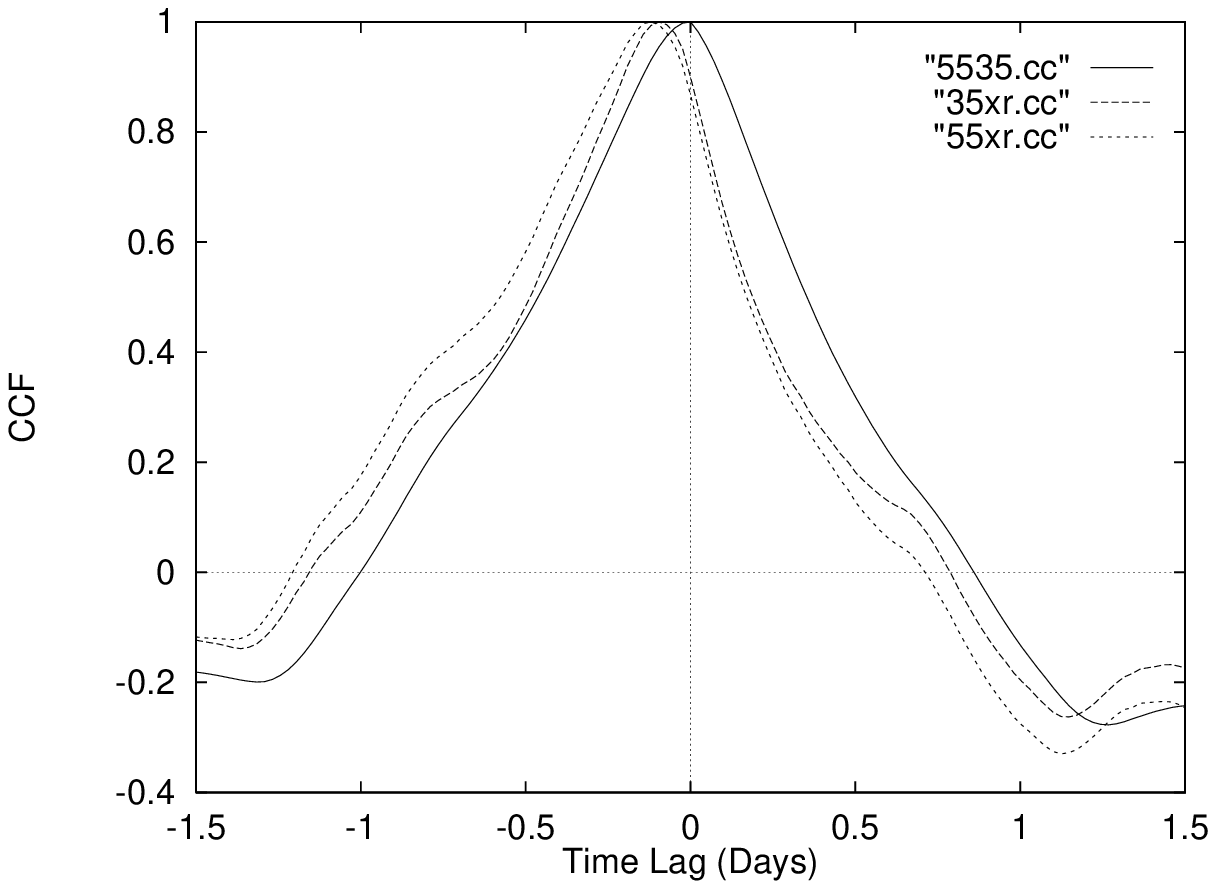,width=.65\textwidth}}
\caption{The Cross Correlation Function between the observed and 
the model light curves of our models for $M = 10^7$ M$_{\odot}$. 
The solid line is the CCF 
between \l\ls 5510 and 3590 \AA. The long-dashed line between 
\ls 3590 \AA~ and the observed X-rays, while the short-dashed line
between \ls 5510 \AA~ and X-rays. Negative lags indicate that the 
second band precedes that of the first. }
\label{fig:m7ccf}
\end{figure*}

\begin{figure*}[hbt]
\centerline{\psfig{file=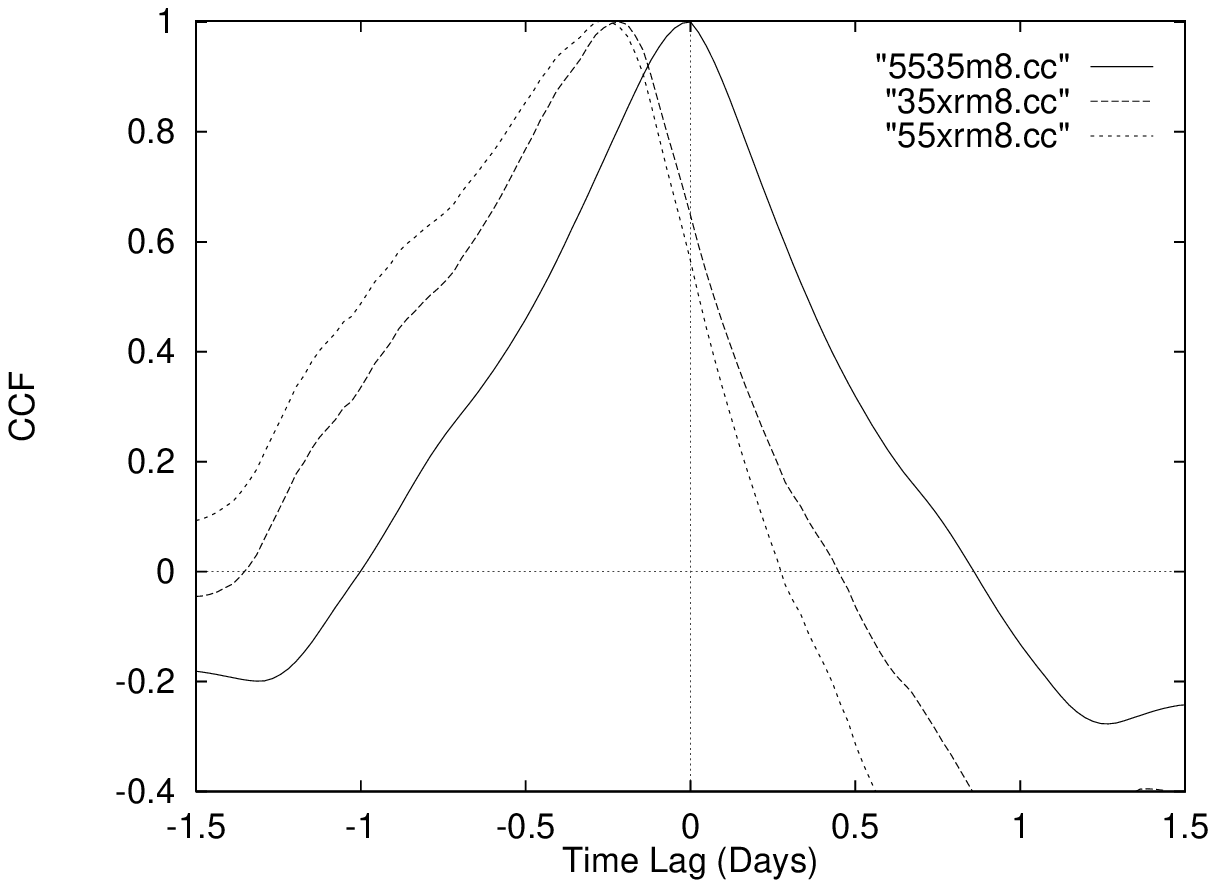,width=.65\textwidth}}
\caption{Same as Figure \ref{fig:m7ccf}  for $M = 10^8$ 
M$_{\odot}$.  }
\label{fig:m8ccf}
\end{figure*}

We quantify further the issue of the lags between the various observed 
bands as a function of the black hole mass by computing their cross 
correlation functions between the model light curves at  \l\ls 
5510 \AA~  and  3590 \AA~ and also between each of the model light
curves at these wavelengths and that of the observed X-rays. These
are presented in figures \ref{fig:m7ccf} and \ref{fig:m8ccf} for 
$M = 10^7,~10^8$ M$_{\odot}$ respectively.

These figures indicate the presence of lags in the cross correlation
functions (CCF) between the observed X-rays and the model 
light curves either wavelength of the reprocessed radiation, 
for either value of 
the black hole mass. However, for $M = 10^7$ M$_{\odot}$ the lags 
are too  small ($\lsim 0.1$ days) to be easily detectable in 
the light curves of figures \ref{fig:m7xrelfl} and \ref{fig:m8xrelfl}. 
This is not the case for $M = 10^8$ M$_{\odot}$,
for which the lag between X-rays and optical emission (of either
wavelength) is apparent and of magnitude $\simeq 0.25$ days, with 
the lags increasing still further for $\theta = 85^{\circ}$. Unfortunately, 
the observed cross correlation functions between the same bands are
much broader than those presented above to allow a direct comparison
(see figure 6 of Edelson et al. 2000); however, there is indication
in the latter that the X-rays preceed the optical emission by 
about 0.25 days.
On the other hand, CCF the model light curves between \l\ls 3590 and 
5510 \AA~ do not indicate 
the presence of any lags between these bands, even for the larger
value of the black hole mass, just as is the case with the 
observed light curves of the same bands. They nonetheless exhibit 
an asymmetry between positive and negative lags suggestive 
of different durations for the reprocessed emission in these
two bands. At this point, one should bear in mind that the 
absence or presence of lags in the  cross correlation function  
depends also on the form of the input signal if there is a large
range of  radii which contribute to the  reprocessed emission. 

\section{The Disk Spectra}

Having obtained a range of values for parameter of the system
from the timing observations and constraints, we use these
to compute the corresponding O--UV and X-ray reflection spectra, which
we then compare to observation. For the O--UV spectra we make the 
usual assumption that the local emission is that of a black body 
which irradiates the locally produced luminosity, as done before 
in the literature. For the X-ray spectra we use the self-consistent 
procedure discussed in NKK and further elaborated in Nayakshin (2000).
This procedure is directly applicable to the generic model
whose properties we attempt to test herein. 
This combined spectro-temporal approach puts far more severe 
and comprehensive restrictions on the specific model than its 
timing or spectral properties alone. We hope that such a combined
approach will eventually point to the direction of class of 
models which can accommodate these  combined constraints.

\subsection{The Optical--UV Model Spectra}

The equations used to compute the time dependent light 
curves to X-ray reprocessing on the disk (in particular 
Eq. (\ref{ouvlcv}) above) can be also used 
for the computation of the spectra of the geometrically 
thin, optically thick disk responsible for the O -- UV 
emission. It has been argued long ago that the spectra of AGN
in this wavelength range are well fit by such models (Malkan \&
Sargent 1983; Malkan 1984; Laor \& Netzer 1989; Sun \& Malkan
1989). 

\begin{figure*}[hbt]
\centerline{\psfig{file=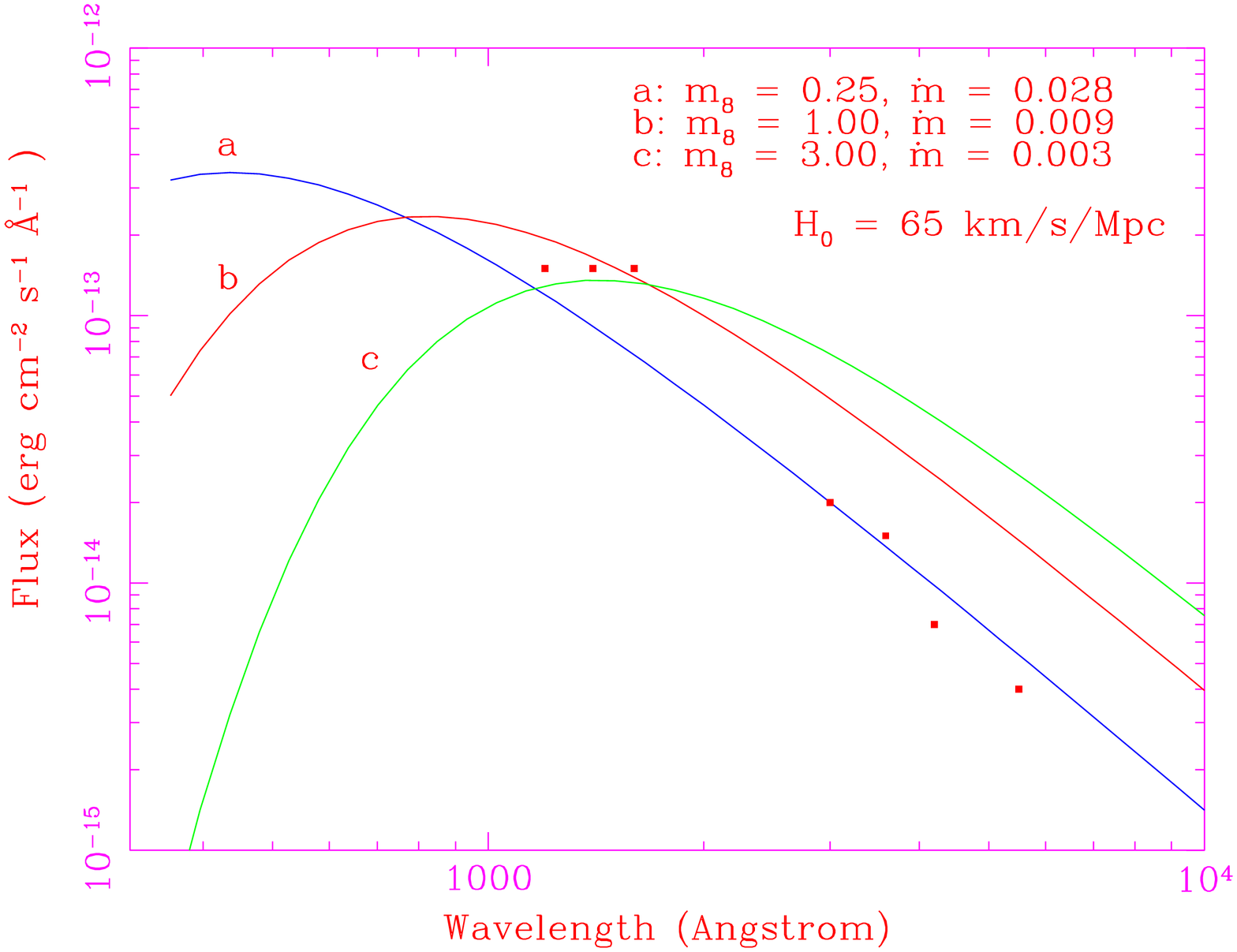,width=.65\textwidth,angle=-90}}
\caption{The Optical -- UV spectra from a multicolor black 
body disk compared to observations. The letters a, b, c 
indicate  the combination of mass (in units of $M = 10^8$ 
solar masses) and accretion rate (in units of the Eddington rate)
used  for producing each  curve. The data, read off from the 
paper of Edelson et al. (2000), are also shown (filled squares)
in the figure for comparison to the models. }
\label{fig:ouvspectra}
\end{figure*}

In order to do so, we  assume that the X-ray flux remains constant
and compute the resulting flux $f_{\lambda}(t)$ for different 
values of the wavelength rather than the time (technically, we 
allow only one step in the $\tau$--integration in Eq. (\ref{ouvlcv}), 
while still integrating the emission over all $R$ for each value
of \l). The values of the black hole mass, $M$, and accretion rate
(in units of the Eddington rate), $\dot m$, are chosen so that they 
provide reasonable eye-ball  fits to the observed fluxes as given in 
Edelson et al. (2000). In order to convert our luminosity values
(derived from the use of a given value of $\dot m$) to observed
flux, the value of the Hubble constant $H_0 = 65 ~\kms$  Mpc$^{-1}$
was used. The results of these calculations are given in Figure
\ref{fig:ouvspectra}.

It is apparent from this figure that no single combination of these
values provides (even) a rough fit to the combined data.
While the values  $M_8 \simeq 3, \dot m \simeq 0.003$ ($\dot m$ is the
accretion rate in units of the Eddington accretion rate) could presumably
provide a reasonable rough fit of the UV data (the flux of this part
of the spectrum appears to be constant), they over produce the optical
luminosity by about a  factor of 5. Decreasing the value of the mass
to $M_8 = 1$, with a concomitant increase in the accretion rate 
to $\dot m = 0.009$ to keep the total luminosity constant, while it 
reduces the discrepancy of the flux in the optical part of the 
spectrum, (the slope of the spectrum actually is quite close to 
that observed), indicates an increase of the UV flux to shorter 
wavelengths which is not observed. Finally, the  combination
$(M_8 = 0.25, \dot m = 0.028)$, provides a reasonable eye ball 
fit to the optical data but fails to reproducing the UV data by a 
a large margin. Similar  combinations with smaller vales of $M_8$, give
spectra which peak at too short wavelengths to provide any useful
fits to the UV data.

\subsection{The X-ray reflection spectra}\label{sect:xray}

We compute the X-ray reflected spectra assuming that the X-ray
spectral index is $\Gamma = 1.5$, as appropriate for NGC~3516 based on
results of Nandra et al. (1999). Further, we use $\eta_X = 0.3$,
exponential cutoff energy $E_{cut} = 100$ keV, and two values for the
BH mass, $10^7$ and $10^8$ M$_{\odot}$.  Using the approach described
in NKK and the associated numerical code, we compute the reflected
spectra at 10 different radii (measured in units of the \sw radius),
$r_k = r_0 \times 2^{(k-1)/2}$ where $r_0 = 3.5$ and $k = 1, 2, ...
10$ and for 10 different viewing angles. We then define a spectrum for
an arbitrary value of $r$ and any viewing angle to be a linear
interpolation of the spectra computed for the nearest values of $r_k$
and $r_{k+1}$ as well as viewing angles. Finally, we integrate over
the accretion disk surface to obtain the full disk spectra at a given
viewing angle, taking into account the gravitational redshift in the
photon energy and Doppler boosting due to Keplerian disk rotation (for
a non-rotating black hole). A more complete discussion of this
procedure is given in Nayakshin (2000b). In addition, the detailed
structure of the illuminated atmosphere of the disk at $r = 6$ for the
lamppost model, and the reflected spectra are presented in Nayakshin
\& Kallman (2000).

\begin{figure*}[t]
\centerline{\psfig{file=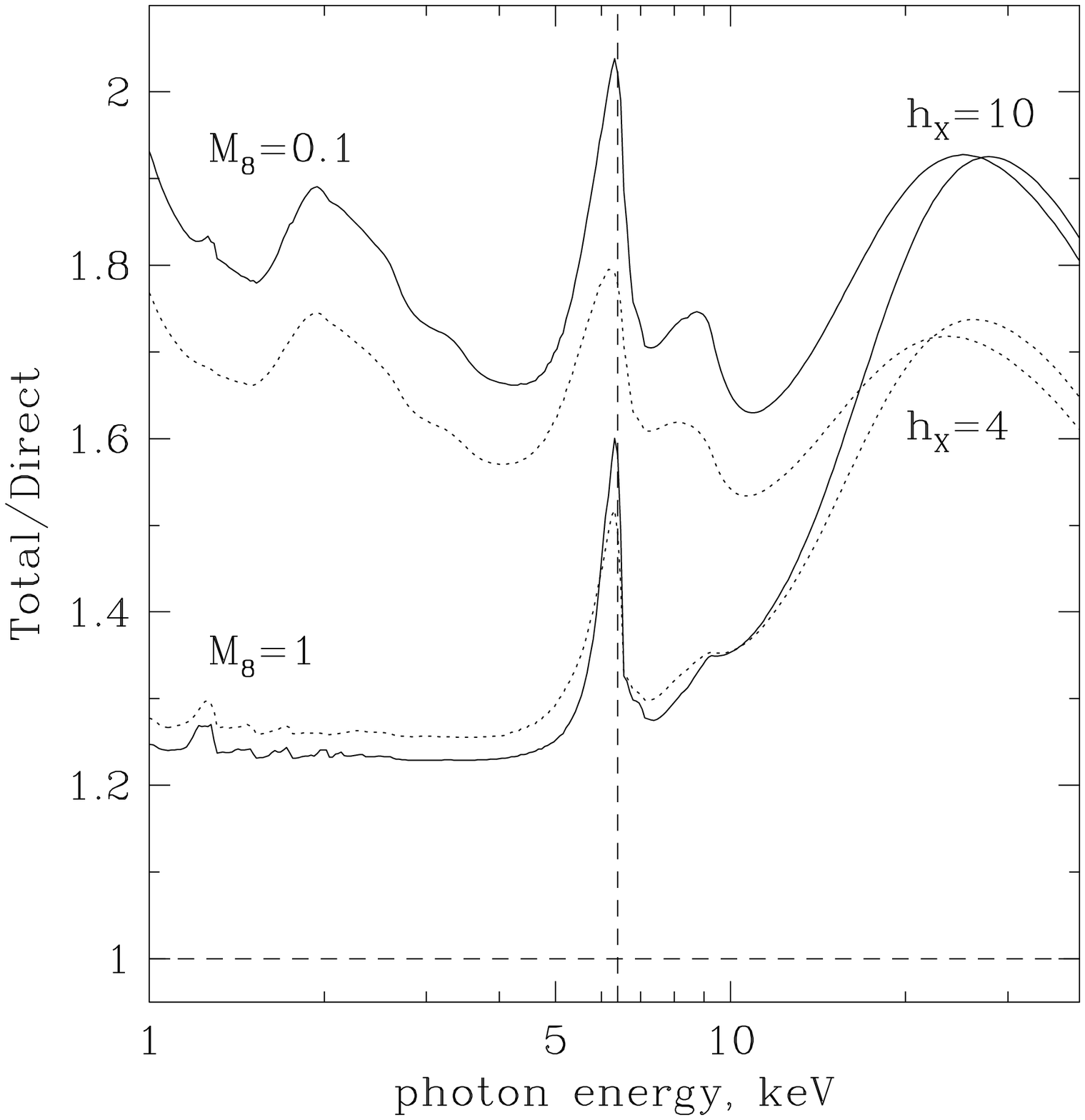,width=.65\textwidth}}
\caption{Full disk spectra for 4 models presented as the ratio of the
total observed spectrum at the inclination angle of $13^{\circ}$ to
the direct X-ray continuum from the lamppost. The solid curves are
computed for the height of the X-ray source of $h_X = 10 R_s$, wheres
the dotted curves are for $h_X = 4 R_s$. The values of the black hole
masses are shown in the Figure. Note that the normalization of the
reflection hump decreases when $h_X$ decreases because X-ray flux
illuminating $R < 3 R_s$ never reflects in our calculations. Also note
strong non-power-law character of the reflected continuum at {\em any}
energy for the smaller black hole mass which comes about because the
X-ray skin is Thomson thick and not completely ionized.}
\label{fig:fullspectra}
\end{figure*}

Figure \ref{fig:fullspectra} shows the reflected spectra at the
inclination angle of $i = 90^{\circ} - \theta =13^{\circ}$ 
(this angle appears to reproduce the Nandra et al. 1999 Fe line 
profile best, see below).  More precisely, the curves in this 
figure exhibit the ratio of the observed spectrum (i.e Source + 
Reflected) with the effects of the observer viewing angle in 
full consideration, to that of the Source alone (assumed to be 
a power law), much in the spirit the ratios of data-to-model  
are presented by the observers. One should note that because 
the highly ionized skin reflects a large fraction of the photons 
even at soft X-ray energies, the ratio of the total observed 
spectrum to the direct continuum X-ray emission can be 
substantially higher than unity for all photon energies (see
NKK). [Therefore, an observer fitting the spectrum with a 
Power law + Reflection from {\sl neutral} material (which reflects
very little at energies $E \lsim 5$ keV) would severely 
underestimate the reflection fraction and therefore the 
solid angle of the source subtended by the reprocessing 
``cold" matter (Done \& Nayakshin 2000)]. 
Figure \ref{fig:fullspectra} demonstrates also the obvious fact 
that  a decrease in the height of the X-ray source leads to a smaller
fraction of $L_X$ intercepted by the disk and therefore to 
less reflected continuum and the \fe emission (dashed lines). 

The spectra corresponding to the black hole mass value of $M_8=0.1$
are rather unusual in the sense that they deviate very strongly from
the roughly power-law type spectra, typical of Seyfert 1 nuclei in the
$\sim$ 1--15 keV range.  For this value of the mass and the observed
X-ray luminosity, the ionized skin of the irradiated disk is
Thomson thick but not completely ionized (i.e. ``warm").  This leads
to an enormous absorption edge (blended with Fe recombination
continuum), and a very strong emission feature between $\sim 1.5$ and
4 keV, which is due to a blend of recombination continua from elements
Mg, Si and S. As discussed in detail in Nayakshin \& Kallman (2000),
these features, as well as the He-like Fe line at $\sim 6.7$ keV
(further broadened and shifted by Compton scattering and relativistic
effects), are due to the presence of this ``warm'' skin on the disk
surface. The absence of these features in the spectrum of NGC 3516,
and in the spectra of Seyfert 1 AGN in general, seems to argue against
this particular range of parameters for this specific X-ray reflection
model.

For the larger value of the mass, $M_8=1$, the X-ray flux on the
accretion disk is smaller, leading to a correspondingly thinner
ionized skin. As a result, even though the skin is still not
completely ionized and does include emission from the transitions
discussed above, its relative contribution to the reflected spectra is
small compared to that of the underlying, neutral, cold layer. 
Therefore the spectra are much more similar to those observed
that are generally fit with reflection from ``cold", neutral matter e.g.,
Basko et al. 1974; Lightman \&1988; Magdziarz \& Zdziarski 1995;
Poutanen, Nagendra \& Svensson 1996). The fits of the
continuum-reflection spectra alone, therefore, seem to favor this
larger value for the black hole mass, in sharp contrast to synchrony
in the O--UV variations which is more consistent with the smaller
($M_8 = 0.1$) value.  Let us now make a detailed
comparison\footnote{Note that a study of the origin of the
``absorption'' feature reported by these authors is beyond the scope
of this paper at this time (but see Ruszkowski \& Fabian 2000)} of the
4 computed spectra with that observed by Nandra et al. (1999).

\subsection{The Fe Line Profiles.}\label{sect:feline}

\begin{figure*}[h]
\centerline{\psfig{file=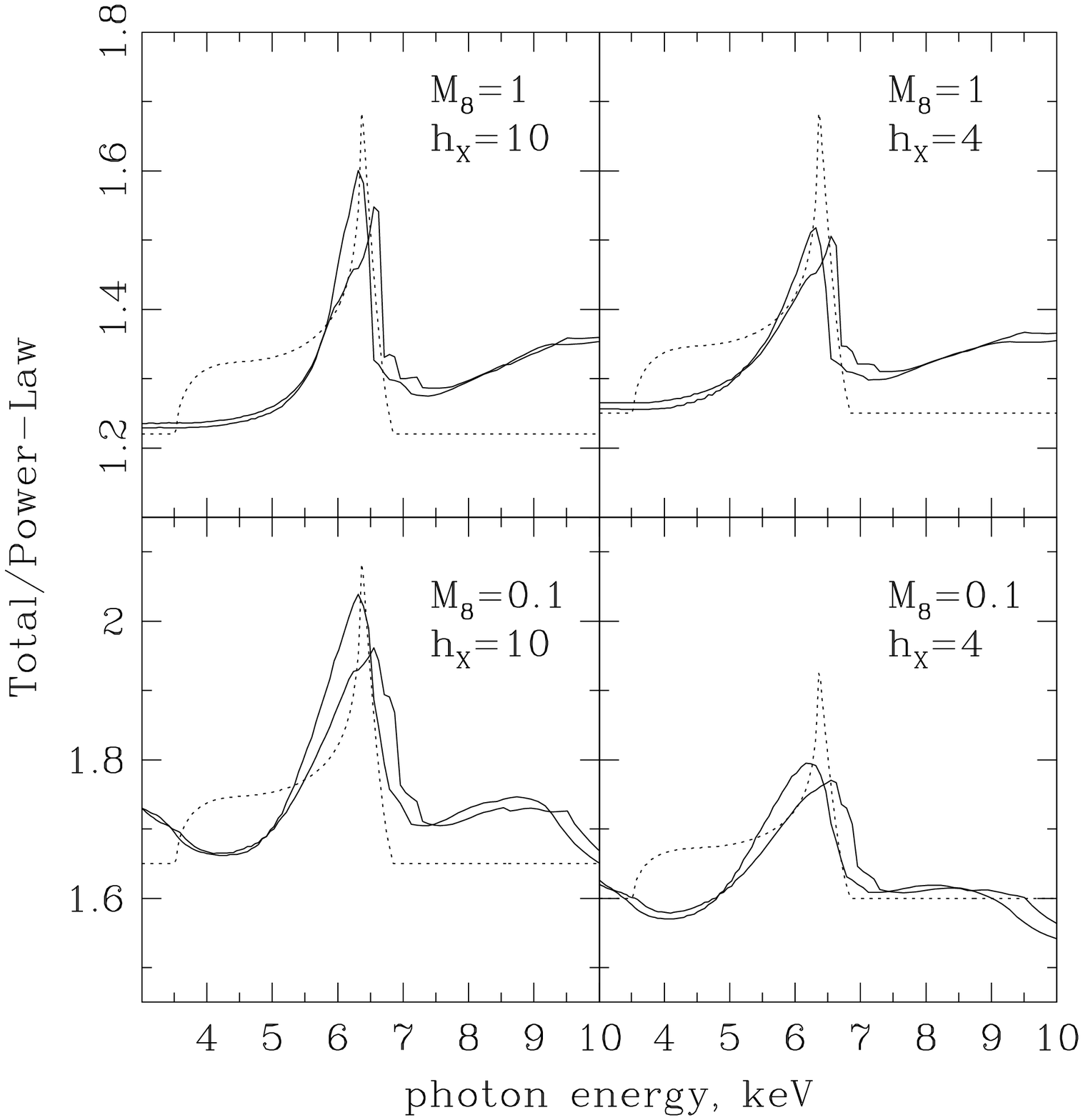,width=.8\textwidth}}
\caption{Solid curves show same spectra as shown in Figure
\ref{fig:fullspectra}, but for $i =13^{\circ}$ and $26^{\circ}$,
and in the region of 3 -- 10 keV as in Nandra et al. (1999). The
dotted curve shows the best disk-line model fit of Nandra et
al. (1999) to their data, arbitrarily rescaled to provide
approximately same Fe line flux as our curves. Note that in all
cases our models have insufficient amount of line flux at $E \lsim
5.5$ keV.}
\label{fig:4spectra}
\end{figure*}

Much of the significance of the Fe line observations lies in the fact
that they are (occasionally) very broad (usually to the red),
indicating emission from X-ray reprocessing on cold matter in the
black hole vicinity. The Fe line in the spectra of NGC~3516, obtained
during the campaign analyzed herein, was indeed broad (Nandra et
al. 1999), thereby providing additional constraints on the parameters
of this system. In the absence of data associated with the Fe line
variability, we have attempted to delineate these constraints through
the modeling of the line profiles. Note that the line profile is
calculated as a part of our full calculation of the X-ray
illumination, so that the spectra shown below are the same
as those shown of Figure \ref{fig:fullspectra}, except that we 
``zoom-in"  the $2-10$ keV energy region.  In Figure \ref{fig:4spectra}
we present the line profiles so computed for two different values of
the inclination angle namely $i = 13^{\circ}, ~26^{\circ}$ (the two
solid lines; the lowest energy one corresponds to $i = 13^{\circ}$)
and for two different values of the black hole mass ($M_8 = 0.1,~ 1$)
and X-ray source height ($h_X = 4,~ 10$), along with the best fit to
the data by Nandra et al. (1999) (dotted line).  The best fit 
line profile is rescaled so that its flux matches that of our
model disk spectra at $E\sim 3$ keV, while its integrated flux
matches that of our models. The choice of inclination angles used by Nandra
et al. (1999) and ourselves is dictated by the demand that these two
effects conspire to produce a line peak around 6.4 keV, as observed.

It is apparent in the figure that none of our line profiles looks even
approximately similar to that which best fits the data (dotted
line). The main reason is the excessive flux at $E \lsim 5.5$ keV
relative to that at $5.5 \lsim E \lsim 7$ keV in the observed
profile. Indeed, in order to provide a fit to the red wing of the
line, Nandra et al. (1999) used a very steep law for the X-ray
illumination of the disk ($F_X \propto R^{-8}$).  This limits the
emission to a section of the disk located very close to the black
hole, resulting in the observed broad profile.  We have attempted to
simulate this by reducing the X-ray source height $h_X$. While the
resulting line profiles are indeed broader, they still fall far short
from matching those of Nandra et al. (1999).  This is due to both
geometry, which dictates that a larger fraction of $L_X$ be
``waisted'' in the hole below at $R < 3 R_S$, and also to ionization
physics: because the illuminating X-ray flux at small $R$ is now
larger than it is for $h_X = 10 R_S$, the skin is more ionized and
produces less of the He-like Fe line, leading to an overall decrease
of its equivalent width.  We do not think that any value of $h_X$
could produce the required illumination law, at least for a \sw
geometry. Consideration of a Kerr geometry may help in this respect,
however, our models cannot correctly compute the resulting profile
without incorporating the additional details involving the photon
propagation in this geometry. Another possible remedy would be to take
into account the \fe line emission from the cold infalling material
within $R < 3 R_S$ (e.g., Reynolds \& Begelman 1997).  Nonetheless, it
is our contention that concentrating the X-ray flux to a small area in
the hole's vicinity would result in a much smaller variability
amplitude in UV--O wavelengths than observed (see for example
Fig. \ref{fig:m8hxth}), because the UV--O emission comes from $ R \gg
R_S$ where the X-ray flux would be too small.

In relation to our detailed line profiles shown in Fig. \ref{fig:4spectra}
we would like to note the following: For the smaller value of the 
black hole mass, favored by the synchrony in interband variations
($M = 10^7 \msun$),  the reflected spectra are distinctively 
non power-law like, due to the high Thomson depth of the skin. 
There is a drop in the computed spectrum above 
$\sim 9$ keV not observed in the actual data (see Figure 1 in 
Nandra et al. 1999);  the data also show a flux decrease below 
$\sim 4$ keV compared to an increase in the same energy range 
produced by our models. In addition, the line flux contains 
contributions from 6.7 keV (He-like) and 6.9 keV (H-like) ions, 
which increase its width near the line core by an 
amount larger than  allowed by the data\footnote{Out of curiosity, 
we computed a full disk spectrum for $M_8 =0.1$ with 
the Doppler effects turned-off and found the line 
width to be $\sim 1$ keV due to Compton scattering 
in the skin and the presence of He- and H-like ion 
contributions.}. These spectral discrepancies between 
model and data argue convincingly against this  value for the 
black hole mass within the lamp-post model. The spectra for the 
larger mass models ($M_8 = 1$), are significantly different
for the reasons discussed in \S \ref{sect:xray}. The smaller 
Thomson depth of the skin leads to a narrower line, dominated
by the 6.4 keV transitions and broadened by the kinematics
of the  disk. The Fe absorption edge is neutral-like at $\sim 
7.1$ keV, but it is not well seen in the model spectra due to its
blending with the Fe line emission by the combined effects of 
Doppler and gravitational smearing. 
There is no noticeable edge/recombination feature at $\sim 9$ keV,
and also no low energy (at $\simeq 2-4$ keV) Mg, Si and S features 
in agreement with observation (Nandra et al. 1999). The EW of 
the line is a little low, a fact that could be remedied by 
by allowing for a super-solar Fe abundance (see George \& Fabian 1991).
   
the spectra in the region of $\sim$ few keV because there is not
enough column density in completely ionized . Thus, the
reflected spectra are much more power-law like and are not in conflict
with the Nandra et al. (1999) data, unlike $10^7\msun$ case. The
centroid energy of the line in the rest frame of the disk material is
6.4 keV, as observed.

\section{Summary, Conclusions}

We have examined in detail the spectro-temporal properties of the
``standard" model of the central regions in AGN, which consists of a
geometrically thin, optically thick accretion disk along with an
overlying illuminating X-ray source at a height $h_X$, in view of the
observations of the recent monitoring campaign of NGC 3516 (Edelson et
al. 2000). To this end we have produced model light curves at the
observed wavelengths \l\ls 1360, 3590 and 5510 \AA, assuming that they
result from the reprocessing of the observed X-ray flux by the
geometrically thin, optically thick accretion disk.  Comparison of the
timing characteristics (amplitudes, lags between the various
wavelengths) of our model light curves to those observed set the range
of acceptable values for the parameters which characterize geometry of
the model, namely $M, ~\dot m, ~h_X$. Using values of these parameters
derived from the timing considerations, we produced the O--UV
continuum, as well as the detailed X-ray reflection spectra expected
from the X-ray illumination of the accretion disk, taking into
detailed account the ionization, radiative transfer, hydrostatic
balance and kinematic and gravitational effects. The model  light
curves also include (in addition to the variable X-ray reprocessed
flux) a constant contribution due to the energy liberation in the thin
disk itself as inferred from observations of NGC~3516.

Our results can be summarized as follows:

1. The variability amplitudes of the reprocessed emission at
\l\ls 3590, 5510 \AA~ are in reasonable agreement with those 
observed, given the inferred ratio $\eta_X =0.3$ of the 
X-ray-to-disk luminosities and assuming $h_X \simeq 10 R_S$. 
The reprocessed light curves
in these two wavelengths appear to vary in synchrony for 
the smaller of the values of the black hole mass examined 
($M_8 = 0.1$), with only a small lag ($\simeq 0.1$ days) between
the X-rays and the reprocessed radiation, which could be
accommodated by the data. However, the data exhibit a noticeable 
absence of direct correspondence between the X-ray and the 
optical emission, not observed in the model light curves. 
In particular, the X-rays exhibit, at day 916.8, a specific 
large amplitude ($\sim 50\%$) ``flare", of 0.7 day duration 
not observed at \l\ls 3590, 
5510 \AA, while it is very clearly present in the model light 
curves of the same wavelengths. This suggests that the size of 
the reprocessing region should be larger than 0.7 light days. 
Indeed, increasing the black hole mass results in model light 
curves which progressively suppress this feature but which, 
at the same time,  destroy the overall synchrony of variation
between the X-rays and the flux at \ls 3590 and \ls 5510. 
Already for a mass $M_8 = 1$ the lags between the X-rays 
and the model light curves are inconsistent with observation, 
while this feature is still clearly present in the model light 
curve. A mass of $M_8 = 3$ suppresses this feature to the extent 
that it is consistent with observations, but only for \ls 5510 \AA,
thereby not resolving the inconsistency (for brevity we do not
exhibit these results in a figure). Interestingly, however, the 
synchrony in the variation between \l\ls 3590 and 5510 \AA~ is not
destroyed; their CCF still peaks close to zero lag but
it becomes progressively asymmetric with the increase in $M$, more 
than allowed by the data. We have not explored the effects of any
larger values for the mass as they appear unreasonable and
inconsistent with the O--UV continuum  data (see next point). 
One could decrease the X-ray - O--UV lags by increasing the disk
inclination angle, $i$, but this is not allowed by the line profile
data (see last point). We can conclude, therefore, that the 
timing observations alone are not consistent with range of the
model parameters explored herein, and likely with any set of 
values of these parameters. 

2. The  observed O--UV spectral distribution of NGC 3516 is 
generally narrower than that of the ``standard" accretion 
disks. Hence, a reasonable fit of the UV flux ($M_8 \simeq 3$)
overproduces the optical emission, while producing a good fit to 
the optical flux ($M_8 \simeq 0.1$) places the peak emission at 
wavelengths
far shorter of the observed 1200--1500 \AA, underproducing the 
observed flux at this range. Interestingly, each mass range is
consistent with only one specific aspect of the timing data:
the smaller mass with the X-ray - optical variation synchrony
and the larger one with the suppression of the X-ray flux feature 
near day 916.8, in the model light curves.

3. The model X-ray reflection spectra are clearly inconsistent with
those observed for the ``small''  values of the black hole mass 
($M_8 \simeq 0.1$), due to the high Thomson thickness of the non
fully ionized skin, which produces atomic transition features 
not observed in the data. Increasing the mass to $M_8 \simeq 1$ 
leads to a much smaller Thomson depth of the skin, and hence X-ray
reflection spectra resembling those observed (but see the last point).
Note, however, that high black hole masses are strongly ruled out by the
timing data (first point). Significantly, our models predict specific
correlations (hard lags) between photons at $E \simeq 1$ keV and at $E
\simeq 15$ keV, of order $\Delta t \sim h_X/c \sim 10^4 \, M_8$ sec
(for $h_X =10$, as assumed in our models) in the cross spectra of
these energies, even without any lag present in their CCF. The X-ray
sampling rate of the source is sufficiently high that a search for
these lags may be reasonable and possible.

4. The Fe line (more accurately the Fe line complex) observations, in 
conjunction with our models, serve mainly to constrain the inclination 
angle of the disk to $i = 13^{\circ} - 26^{\circ}$. While 
our models produce reasonable equivalent widths for this line,
they are unable to match, even approximately, the best fit of the
observed profile. The main reason is the large red wing of the \fe
line extending to
$E \sim 4$ keV, which requires that most of this emission, and 
therefore the associated X-ray illumination, be confined to a region
very close to the black hole horizon. Our models, with the source 
located at any source height, fall far short from achieving this. 
An extreme Kerr hole would 
help in this respect, however, quantitative models  do not  exist 
for this case; furthermore, it is not apparent how  such 
models would provide a resolution of the issues raised in point 1
above. Finally, concentrating too much of the X-ray luminosity in
the $R\sim$ few $R_S$ region will leave too little of this luminosity
to illuminate the region $R\gsim$ tens of $R_S$ (that emits the O--UV
part of the spectra) so that it will underproduce the amount of the O--UV
variations compared to those observed.

Where do all these leave us? It is apparent that one cannot fulfill all 
the observational constraints within this picture of the 
central engine of AGN. The alternatives are few: (a) Ignore some of 
the constraints as less important and attempt concordance within 
this less restricted constraint list (b) Abandoned this model
 in favor of a different one. 

Of the present constraints, those set by timing seem to be the more 
significant ones, in that by themselves point to a possible internal 
inconsistency of the model. As argued elsewhere
(e.g. Kazanas, Hua \& Titarchuk 1996), spectral fits and observations
can, generally, yield information only about optical depths and 
column densities and that timing information is necessary to 
convert these to the length and densities necessary to determine 
the geometry of the source. The problem with the timing
observations is that it is not known, in the absence of repeated 
observations, whether they reveal a typical or a transient property
of the object under consideration. Clearly ``more observations are
needed", with as broad wavelength coverage as possible, 
given that it is the interband variability which 
provides the most stringent constraints on the models. 

Of our model spectra and their fits to the data, those of the
 O--UV continuum maybe thought as the more suspect given
the simplicity of their assumptions and the possibility of 
their contamination by the host galaxy. However, since that
such a contamination would more likely contribute to the 
optical rather than the UV flux, its removal it would only increase the
discrepancy between model and observation. The X-ray reflection 
spectra, on the other hand, are very robust within the specific 
model examined and seem to argue for the larger values of the mass 
considered. The energy dependent albedo and the related timing 
analysis within the X-ray band itself, if feasible, should clearly 
set additional, independent limits on size of the system. We hope 
that such analysis in this and other objects be carried out and set 
as one of the observing goals.

We would like to thank K. Nandra for providing us with the X-ray 
light curve of NGC 3516.

{}

\end{document}